\documentclass[a4paper, amsfonts, amssymb, amsmath, reprint, showkeys, nofootinbib, longbibliography, twoside]{revtex4-2}
\usepackage[english]{babel}
\usepackage[utf8]{inputenc}
\usepackage{graphicx}
\usepackage{color}
\usepackage{float}

\usepackage{chngcntr}

\usepackage[pdftex,pdftitle={Article},pdfauthor={Author},hidelinks]{hyperref}
\usepackage[T2A,T1]{fontenc}
\usepackage{lipsum} 
\usepackage{soul}
\usepackage[dvipsnames]{xcolor}
\usepackage{dsfont}
\usepackage{siunitx}	    

\usepackage{amsmath}
\usepackage{mathrsfs}
\usepackage{bbm}
\usepackage{amssymb}
\usepackage{bm}             
\usepackage{mathtools}      
\usepackage{cancel}         
\usepackage{tensor}         
\usepackage{subcaption} 
\usepackage{caption}    
\usepackage{ragged2e}

\usepackage{fancyhdr}
\newcommand{\vb}[1]{\boldsymbol{\mathbf{#1}}}

\newcommand{\ad}[1]{{\color{black}#1}}

\newcommand{\sg}[1]{{\color{black}#1}}
\newcommand{\half}{\frac{1}{2}}

\newcommand{\tm}[1]{{\color{black}#1}}

\begin{document}

\title{Anti-hyperuniform Critical States of Active Topological Defects}


\author{Simon G. Andersen$^1$}
\author{Tianxiang Ma$^1$}
\author{Makito F. Katsume$^1$}
\author{Kexin Li$^2$}
\author{Xiao Liu$^2$}
\author{Martin Cramer Pedersen$^1$}
\author{Amin Doostmohammadi$^1$}
\email[Correspondence email address: ]{doostmohammadi@nbi.ku.dk}
\affiliation{$^1$Niels Bohr Institute, University of Copenhagen, Blegdamsvej 17, Copenhagen, Denmark}
\affiliation{$^2$Key Laboratory of Biomechanics and Mechanobiology (Beihang University), Ministry of Education, Beijing Advanced Innovation Center for Biomedical Engineering, School of Biological Science and Medical Engineering, Beihang University, Beijing 100083, China}
\date{\today}

\begin{abstract}

\vspace{2em}
\noindent
Topological defects are fundamental to the collective dynamics of non-equilibrium systems and in active matter, mediating spontaneous flows, dynamic self-organization, and emergent pattern formation.
Here, we reveal critical states in active nematics, marked by slowed defect density relaxation, amplified fluctuations, and heightened sensitivity to activity. Near criticality, defect interactions become long-ranged, scaling with system size, and the system enters an anti-hyperuniform regime with giant number fluctuations of topological defects and defect clustering. This transition reflects a dual scaling behavior: fluctuations are uniform at small scales but become anti-hyperuniform at larger scales, \tm{as supported by experimental measurements on large-field-of-view endothelial monolayers. We find that these anti-hyperuniform states with multiscale defect density fluctuations are robust to varying parameters, introducing frictional damping, and changing boundary conditions.}  Finally, we show that the observed anti-hyperuniformity originates from defect clustering, distinguishing this transition from defect-unbinding or phase separation processes. Beyond fundamental implications for non-equilibrium systems, these results may inform biological contexts where topological defects are integral to processes such as morphogenesis and collective cellular self-organization.
\end{abstract}

\maketitle

\section{Introduction} \label{sec:intro}

\noindent
Active matter is crucial for understanding the collective behavior and emergent phenomena of biological many-body systems, spanning length scales from molecules to animal flocks. Modeling these systems, where each particle expends energy to generate motion and exert forces on its surroundings, has revealed how energy consumption at the microscopic level can drive large-scale patterns and behaviors, and has provided deep insights into the non-equilibrium statistical physics of many-body systems~\cite{ramaswamy_mechanics_2010,marchetti_hydrodynamics_2013,gompper_2020_2020}.

In particular, active nematic models have been successful in describing the collective behavior of eukaryotic and bacterial cells \cite{doostmohammadi_active_2018}.
When hydrodynamic interactions are present,
the ordered nematic state is destabilized by particle activity~\cite{aditi_simha_hydrodynamic_2002}, ultimately leading to active turbulence, and at sufficiently high activity, the creation of topological defects becomes energetically favorable \cite{thampi_active_2016, yeomans_hydrodynamics_2016}.
Such defects have turned out to serve important roles in the collective behavior of living cells, 
notably, in relation to tissue morphogenesis \cite{maroudas–sacks_topological_2021, hoffmann_defect_morphogenesis_2022}, cellular dynamics and organization \cite{bonhoeffer_iso–orientation_1991,saw_topological_2017, copenhagen_topological_2021, doostmohammadi_defect–mediated_2016}, and collective migration \cite{kawaguchi_topological_2017, serra_defect–mediated_2023, yaman_emergence_2019,doostmohammadi_physics_2021}. 

With the growing recognition of the importance of topological defects and their roles in various non-equilibrium systems, it is natural to search for generic patterns of defect organization in active materials. This is important because varying activity levels can significantly alter the dynamics and interactions of active particles, leading to different organizational states and behaviors~\cite{marchetti_hydrodynamics_2013}. 
Moreover, while various methods to control and order topological defects have been introduced both experimentally and in theoretical models~\cite{doostmohammadi_active_2018}, 
the existence of 
generic critical behavior and phase transitions in topological defect dynamics remains an open question.
The presence of criticality in defect organization is particularly intriguing because it can highlight universal principles that govern the behavior of active matter. Critical points often signify a balance between competing forces or interactions, leading to unique and sometimes unexpected phenomena. Investigating these aspects could provide a better understanding of the fundamental principles governing active matter and enhance our ability to manipulate these systems for practical applications.

Here, using large-scale numerical simulations of active nematics we reveal a critical activity threshold that separates distinct states of defect organization. 
This critical point exhibits all the features of a phase transition, including critical slowing down, diverging spatial and temporal correlations, and a sharp peak in the amplitude and lifetime of fluctuations around their steady state means. We further show that as the system approaches the critical point, defect configurations transition from a uniform state, to an anti-uniform state, with anti-hyperuniformity (amplified density fluctuations) peaking at the critical point. At this point, we also estimate the anomalous dimension of the pair correlation function.

We explain the origin of hyperfluctuations in defect density as a self-enhancing nucleation process that favors the formation of new topological defects near existing ones, leading to a clustering effect near the critical activity. Notably, we show that this activity is distinct from the activity thresholds for defect nucleation and unbinding, and that the gap between them remains independent of the system size. \ad{We further show that the observed anti-hyperuniformity and multiscale defect density fluctuations are robust to changing elastic constant, flow aligning, frictional damping, and changing boundary conditions. Finally, we show experimental evidence of the anti-hyperuniformity and multiscale defect density fluctuations in endothelial monolayers.}






\section{Governing Equations} \label{sec:continuum}

\noindent
We employ the well-established equations of active nematohydrodynamics~\cite{doostmohammadi_active_2018}. In this framework, the magnitude and direction of orientational order is contained in the traceless and symmetric nematic order parameter $Q_{ij} = \frac{d q}{d-1}\left(n_i n_j - \frac{\delta_{ij} }{d}\right)$, where $q$ is the magnitude of nematic order, $n_i$ is the director field, \ad{and $d$ is the dimensionality of the system}. 
The field equations are obtained by incorporating activity into the Beris-Edwards equations \cite{beris_thermodynamics_1994}, which determine the evolution of $Q_{ij}$ and the incompressible velocity field $u_i$ through
\begin{subequations}\label{eq : gov-equations}
\begin{align}
(\partial_t + u_k \partial_k) Q_{ij} - S_{ij} &= \Gamma H_{ij},
        \label{eq:beris-ed} \\
    \partial_i u_i &= 0,
        \label{eq:stokes_cont} \\
    \rho (\partial_t + u_k \partial_k)u_i &= \partial_j \Pi_{ij}.
     \label{eq:stokes_full}      
\end{align}
\end{subequations}
The evolution of $Q_{ij}$ is described by Eq. \eqref{eq:beris-ed}, in which the molecular field $H_{ij} = - \frac{\delta F}{\delta Q_{ij}} + \frac{\delta_{ij}}{d} \textup{Tr} \frac{\delta F}{\delta Q_{kl}}$ determines the relaxation of $Q_{ij}$ towards the minimum of the free energy $F = \int d^dx~ [\mathcal{A}\left(1 - Q_{ij}Q_{ji}\right)^2 + \frac{K}{2} (\partial_k Q_{ij})^2]$. The first term in $F$ is the Landau-de Gennes free energy with coefficient $\mathcal{A}$, written such that the ground state of $Q_{ij}$ is the perfectly ordered nematic state \cite{yeomans_hydrodynamics_2016}. The elastic energy penalties of bend and splay distortions in the director field are assumed to be equal, thus allowing a single Frank elastic constant $K$ to account for both \cite{doostmohammadi_active_2018}. 

The shape of the nematogens cause them to rotate in response to velocity gradients in the flow field, as is accounted for through the co-rotational advection term $S_{ij} = (\lambda E_{ik} + \Omega_{ik}) (Q_{kj} + \frac{\delta_{kj}}{d}) +  (\lambda E_{kj} - \Omega_{kj}) (Q_{ik} + \frac{\delta_{ik}}{d}) - 2 \lambda  (Q_{ij} + \frac{\delta_{ij}}{d}) (Q_{kl} \partial_k u_l )$, where $E_{ij} = \half (\partial_i u_j + \partial_j u_i)$, $\Omega_{ij} = \half (\partial_j u_i - \partial_i u_j)$ denote the strain rate and vorticity tensors, respectively, and $\lambda$ is the alignment parameter. $\lambda$ determines whether the response of the nematogens is dominated by strain or vorticity, with $\lambda > 0$ corresponding to rod-like particles, and $\lambda < 0$ corresponding to disk-like particles \cite{yeomans_hydrodynamics_2016}. 

The evolution of the velocity field $u_i$ is governed by the generalized Navier-Stokes equations (\eqref{eq:stokes_cont}, \eqref{eq:stokes_full}), where $\Pi_{ij} = \Pi_{ij}^{\textup{viscous}} + \Pi_{ij}^{\textup{elastic}} + \Pi_{ij}^{\textup{active}}$ is a generalized stress tensor that accounts for the additional complications caused by the shape of the nematogens as well as their activity \cite{thampi_active_2016}. 

The viscous contribution has its usual form $\Pi_{ij}^{\textup{viscous}} = 2 \eta E_{ij}$, where $\eta$ is the isotropic viscosity of the fluid. The elastic part consists of a bulk pressure contribution, in addition to a number of terms incorporating the effect of back-flow caused by the rotation of nematogens, $\Pi_{ij}^{\textup{elastic}} = - P \delta_{ij} + 2 \lambda (Q_{ij} + \frac{\delta_{ij}}{d}) Q_{kl} H_{lk} - \lambda H_{ik}  (Q_{kj} + \frac{\delta_{kj}}{d}) - \lambda H_{kj}  (Q_{ik} + \frac{\delta_{ik}}{d}) 
- \partial_i Q_{kl} \frac{\delta F}{\delta \partial_j Q_{lk}} + Q_{ik} H_{kj} - H_{ik} Q_{kj}$. Finally, the activity $\zeta$ is included through $\Pi_{ij}^{\textup{active}} = - \zeta Q_{ij}$ \cite{yeomans_hydrodynamics_2016}.
\begin{figure*}[t!]
    \centering
        \centering
        \includegraphics[width=\textwidth]{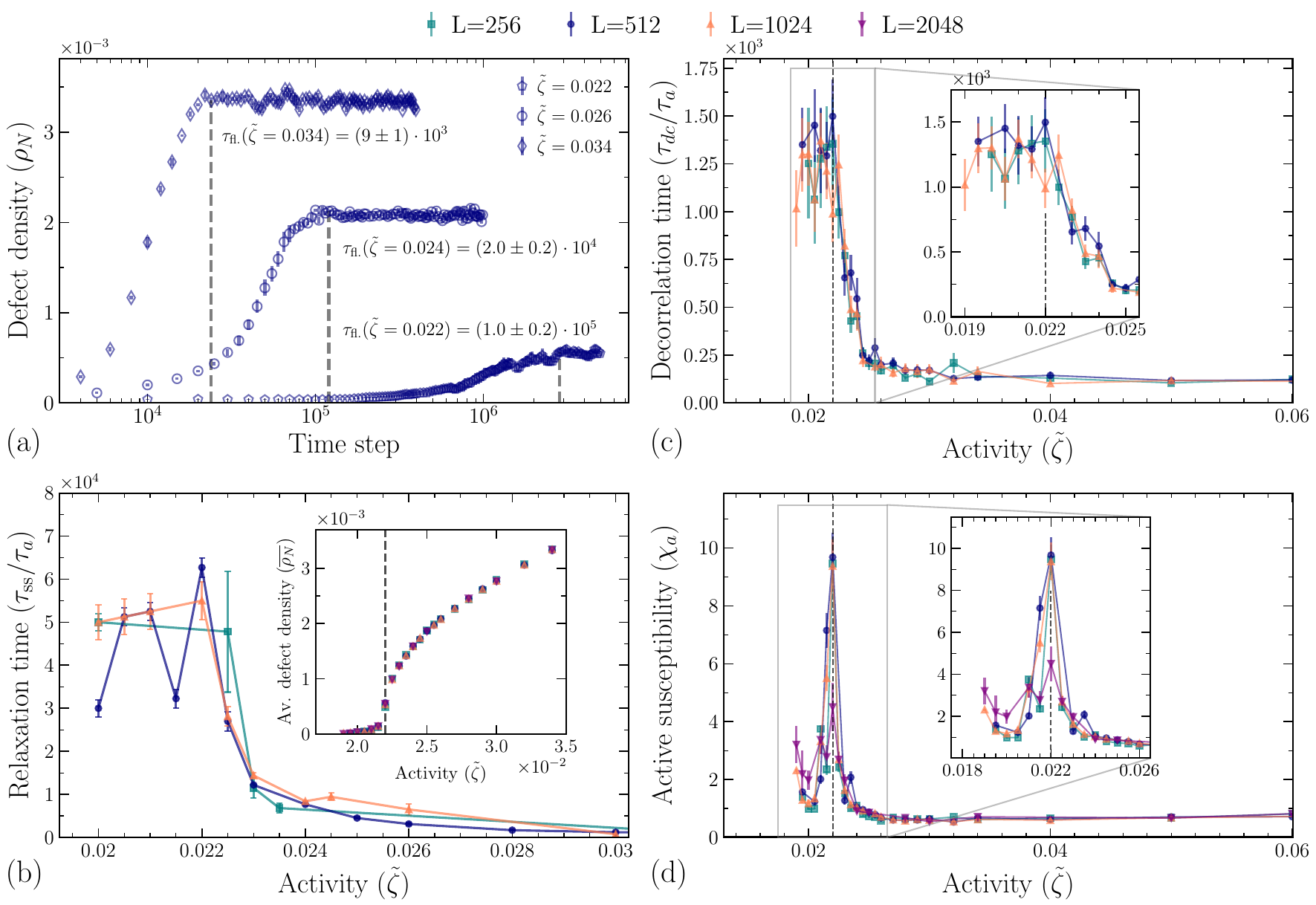} %
    \caption{{\bf Critical slowing down, enhanced temporal decorrelation, and peak in active susceptibility of topological defects at low activity.} (a) The global defect density, $\rho_N$, (averaged over 10 realizations) for $L = 512$ and $\tilde \zeta = \{0.022, 0.026, 0.034 \}$. The relaxation time $\tau_{\text{ss}}$ (indicated by vertical lines) increases by two orders of magnitude as $\tilde \zeta$ is decreased, \sg{while the lifetime of fluctuations $\tau_{\text{fl.}}$ in  $\rho_N$ around its steady state mean $\overline{\rho_N}$ increases by one order of magnitude.} (b) The relaxation time (normalized by the active time scale $\tau_a \sim \eta \tilde \zeta^{-1}$) across activities for 3 system sizes. It increases rapidly as the activity is decreased below $\tilde \zeta \approx 0.024$.
    The inset shows the average global defect density, $\overline{\rho_N}$.
    Near $\tilde \zeta=0.022$ (vertical line), it is highly sensitive to small changes in the activity.
    (c) Decorrelation time $\tau_{\text{dc}}$ (average of realizations) normalized by $\tau_a$. It is roughly one order of magnitude larger for $\tilde \zeta\lessapprox 0.022$ as compared to larger activities. (d) The active susceptibility, $\chi_a =  (\overline {N^2} - \overline  N ^2) / \overline{N}$, where $\overline{N}$ is the average number of defects in a frame, against activity. It is roughly symmetric around its peak at $\tilde \zeta=0.022$.}
    \label{fig:av_dens_superfig}
\end{figure*}

\vspace{1em}
\noindent
Eqs. \eqref{eq : gov-equations} have been solved numerically using the hybrid Lattice Boltzmann method (App. \ref{methods:hb}) for four system sizes and a range of extensile activities, with 5 and 10 realizations of each simulation for $L=2048$ and $L\in \{256,512,1024\}$, respectively. Boundary conditions are periodic, and each simulation has been initialized in an ordered nematic state with slight local perturbations in the director field (App. \ref{methods:sim_params}).

Unless otherwise stated, all parameters are expressed in units of the lattice spacing $\Delta x = 1$, the simulation time step $\Delta t = 1$, and the Landau coefficient $\mathcal{A} = 1$ (Tab. \ref{tab:model_params}). \sg{All quantities can be converted to physical units accordingly depending on the material of interest \cite{cates_shearing2008, thampi_velocity_2013} (for an overview of relevant length and time scales, see \cite{hemingway_correlation_2016}).
The activity parameter is expressed in dimensionless form, $\tilde{\zeta}=\zeta/\mathcal{A}$, which corresponds to the squared ratio of the coherence length $r_c \sim \sqrt{K/\mathcal{A}}$ to the active length scale $r_a \sim \sqrt{K/\zeta}$.}



\section{Results} \label{sec:results}

\subsection{Behavior of Global Defect Density indicates Phase Transition at Low Activity} \label{subsec:defect_dens}

\noindent
We begin by examining the relaxation dynamics of the global defect density in active nematics. Relaxation times provide insight into the timescales, over which systems approach steady states, and can reveal signatures of critical phenomena such as critical slowing down \cite{lubeck_nonequilibrium_2004, tirabassi_2024}.
For each system size with side length $L$, each realization, and for activity range $\tilde \zeta \in [0.019, 0.1]$, we collect samples at steady state, and the relaxation time $\tau_{\text{ss}}$ for the
global defect density $\rho_N(t)$ (averaged across realizations) is estimated as the number of time steps to reach steady state. \sg{Finally, the fluctuation lifetime $\tau_{\textup{fl.}}$ of $\rho_{N}(t)$ around its steady state mean $\overline{\rho_N}$ is estimated as the average time interval between sign changes of the residual
$\rho_N(t) - \overline{\rho_N}$.} (Fig. \ref{fig:av_dens_superfig}a).

As the activity is decreased below $\tilde \zeta \approx 0.024$, the relaxation time $\tau_{\text{ss}}$ increases by two orders of magnitude as compared to higher activities, \sg{while the lifetime of fluctuations $\tau_{\text{fl.}}$ increases by roughly one order of magnitude}. We refer to this activity regime, where $\tilde \zeta <0.024$, as the {\it proliferation regime}. 

\begin{figure*}[t!]
        \centering   \includegraphics[width=\textwidth]{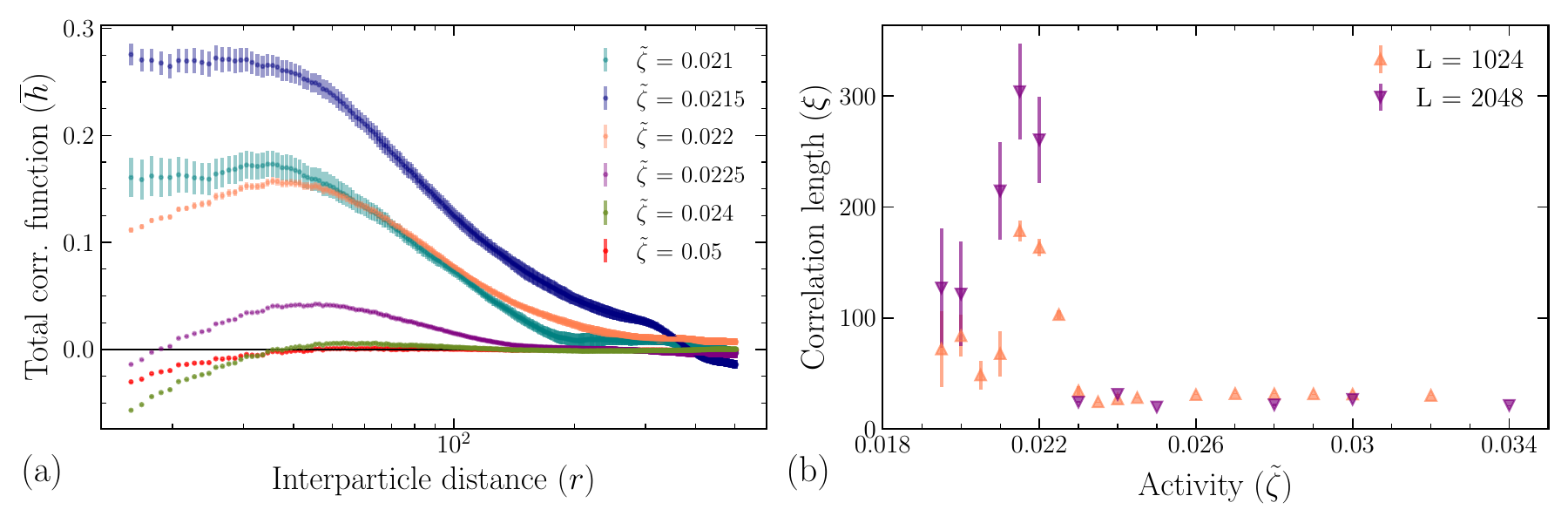} %
    \caption{{\bf Emergence of long-ranged defect-defect pair correlations.} (a) Sample average of the total correlation function $\overline{h}$ for activities near the transition ($L=2048$). For $\tilde \zeta \leq 0.022$, pair correlations are positive and long-ranged.
    For $\tilde \zeta \in [0.0225,0.024]$, anti-correlations at short range give way to positive correlations at long range, whereas only anti-correlations are present 
    for higher activities.
    (b) Estimates of the correlation length $\xi(\tilde \zeta)$ (averaged over realizations) as the smallest $r^*$ satisfying $|\overline{h}(\tilde \zeta, r^*)| - \textup{err}(\overline{h}(\tilde \zeta, r^*)) < 10^{-2}$, \sg{where $\textup{err}(\overline{h}(\tilde \zeta, r))$ is the uncertainty on $\overline{h}(r)$ for activity $\tilde \zeta$}. For both systems, it has a peak for $\tilde \zeta \in  [0.0215,0.022]$.}
    \label{fig:pcf_corr_len}
\end{figure*}

The decreasing trend of the relaxation time with increasing activity aligns with the behavior of the active time scale, which is a measure of the relative strength of viscous and active stresses and scales as $\tau_a \sim \eta /\tilde \zeta$ \cite{doostmohammadi_active_2018}. However, normalizing $\tau_{\text{ss}}$ by $\tau_a$ does not account for the two order of magnitude increase in the relaxation time (Fig. \ref{fig:av_dens_superfig}b). Instead, the relaxation time peaks at $\tilde \zeta \approx 0.022$, suggesting the possibility of critical slowing down as the underlying mechanism.
Near this activity, we also see that the average global defect density $\overline{\rho_N}$ is highly sensitive to small changes in the activity (Fig. \ref{fig:av_dens_superfig}b inset).

To further investigate the hypothesis of critical slowing down, we analyze the temporal correlations in the defect density $\rho_N(t)$. Specifically, we measure the decorrelation time $\tau_{\text{dc}}$ as the smallest time increment $\Delta t^*$, for which the autocorrelation function satisfies $C(\Delta t^*) < 0.2$ (Fig. \ref{fig:av_dens_superfig}c).
 Across different system sizes, the decorrelation time $\tau_{\text{dc}}$ (averaged over realizations) is about one order of magnitude larger in the proliferation regime than at larger activities, with peak values occurring below the same activity threshold ($\tilde \zeta \approx 0.022$) observed for critical slowing down.
 This rapid increase of the decorrelation time as the activity is decreased towards $\tilde \zeta \approx 0.022$ provides further indications of possible critical behavior.

If a critical point exists, fluctuations in physical quantities relative to their steady-state means are expected to get amplified as criticality is approached \cite{lubeck_nonequilibrium_2004}. To quantify the amplitude of such fluctuations in $N$, we define an `active susceptibility', $\chi_{a} = (\langle N^2 \rangle - \langle N \rangle ^2)/{\langle N \rangle}$, as a non-equilibrium analog to the susceptibility of the liquid-gas transition \cite{torquato_hyperuniform_2018}. Across different system sizes, $\chi_a$ is symmetric around its peak at $\tilde \zeta = 0.022$ (Fig. \ref{fig:av_dens_superfig}d), with a width roughly corresponding to that observed for the decorrelation time $\tau_{\text{dc}}$, further supporting the identification of a critical activity threshold in the proliferation regime.

Finally, we observe defects to be unbound whenever present (see supplementary videos 1-3). This is in line with the expectation that, in the absence of screening or frictional damping, any nonzero activity leads to hydrodynamic instabilities and defect unbinding \cite{thampi_instabilities_2014, thampi_2014,thampi_active_2016}.

Taken together, measurements of the relaxation time, decorrelation time, and active susceptibility all point to the existence of an activity threshold governing the behavior of topological defect density in active nematics. Notably, the observed activity threshold $\tilde \zeta = 0.022$, hitherto refereed to as the critical activity $\tilde \zeta_c$, is not only distinct from but also remains well-separated from the activity threshold for defect nucleation and unbinding, with the gap between them remaining constant and independent of system size. \ad{The critical value of the non-dimensional activity in our system reflects a crossover where activity-induced deformations begin to compete significantly with elastic coherence. While the qualitative nature of this transition is robust, the exact threshold value of the dimensionless activity depends on material parameters such as the elastic constant, frictional damping, flow-aligning parameter, magnitude of the nematic order, and may also include numerical prefactors. For this reason, while the location of the transition is meaningful and reproducible within a given system, it should not be interpreted as a universal number.}


\subsection{Emergence of Long-Ranged Defect-Defect Pair Correlations} \label{subsec:pcf}

\noindent
Thus far, we have focused on the global defect density, but understanding the spatial distributions of defects offers deeper insights into the organization of the system. To this end, we examine the isotropic pair correlation function, $g_2(r)$ (App. \ref{methods:pcf}).
\ad{Previous studies measured the pair distribution function $g(r)$ of topological defects in experimental realizations of active nematics, and in simulations~\cite{decamp_orientational_2015,kumar_tunable_2018}, focusing primarily on short-range positional and orientational correlations of $+1/2$ defects in regimes characterized by turbulent or coherent defect motion. Our approach differs in both scope and motivation. We analyze the spatial organization of the full defect population, including both $+1/2$ and $-1/2$ defects, and focus on long-range statistical properties such as number fluctuations and structure factors. This allows us to characterize the emergence of hyperuniform and anti-hyperuniform scaling regimes, and to identify a critical activity at which the nature of defect organization undergoes a qualitative change. Our analysis is thus focused on the collective organization of topological defects, which, in active nematics, inherently involves both $+1/2$ and $-1/2$  defects as co-dependent excitations. These defect pairs emerge and annihilate together to preserve topological charge neutrality, and their mutual interactions, particularly the self-propelled dynamics of $+1/2$ defects and the elastic attraction with $-1/2$ defects, are essential features of the defect-mediated dynamics in these systems~\cite{giomi_defect_2013,pismen_dynamics_2013,thampi_2014}. While it is true that the spatial correlations of $+1/2$ and $-1/2$ defects can individually carry distinct signatures, e.g. due to the polar nature of the $+1/2$ defect, the focus of our analysis is not on species-specific properties, but on the emergent organization of the defect network as a whole. This global viewpoint is particularly relevant for probing long-range correlations and scaling laws (sections B and C of the manuscript), where the interplay between both defect types governs the large-scale structure.} \sg{For completeness, however, we have also shown that the conclusions of this and the following section are true even when $+1/2$ or $-1/2$ defect species are considered separately (Figs. \ref{fig:suppfig:pcf_signed} \& \ref{fig:suppfigsfac_signed})}.

For $\tilde \zeta \leq \tilde \zeta_c = 0.022$, the sample average of the total pair correlation $\overline{h} = \overline{g_2} -1$ is long-ranged and positive, whereas for $\tilde \zeta > 0.024$, it is negative, indicating anti-correlation, and decays rapidly (Fig. \ref{fig:pcf_corr_len}a). Interestingly, for $\tilde \zeta \in [0.0225,0.024]$, the behavior transitions: anti-correlations at short distances give way to positive correlations at longer distances. This suggests a complex interplay between defect interactions near the critical activity threshold. 
\begin{figure*}[t!]
        \centering   \includegraphics[width=\textwidth]{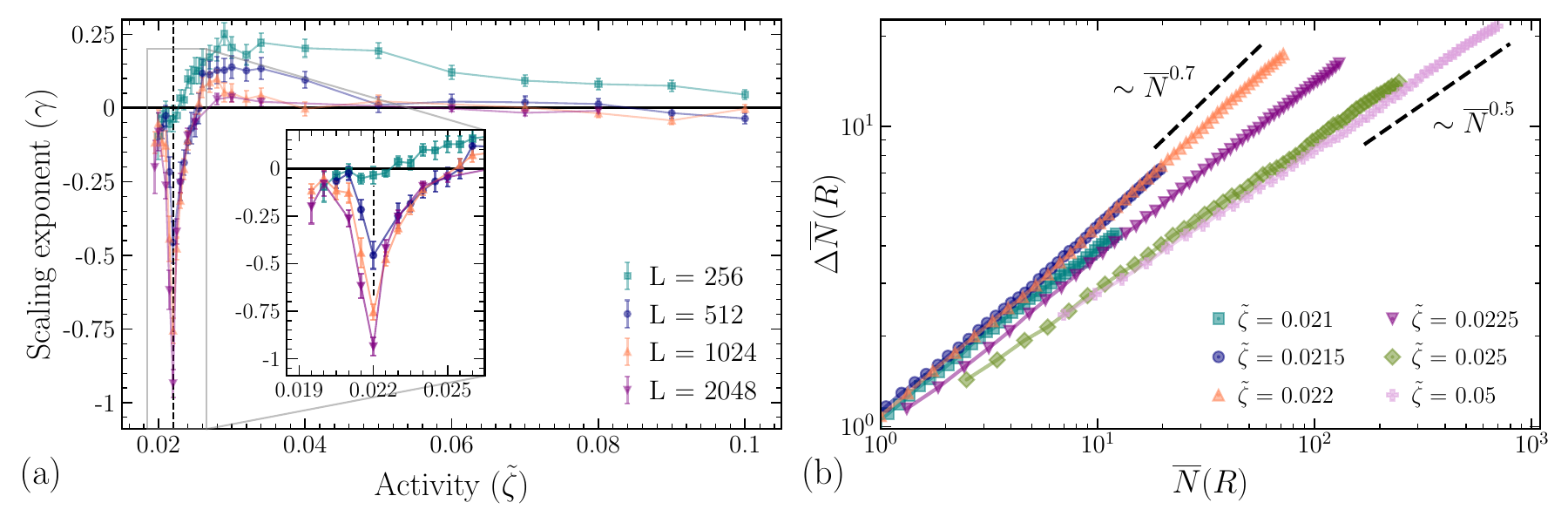} 
    \caption{{\bf Emergence of anti-hyperuniform defect states near the critical point}. (a) The estimated hyperuniformity exponents, $\gamma$, as obtained by fitting the small wavenumber tails of the structure factor (averaged over samples and orientations), $\overline{S}(|\vb{k}|) \sim |\vb{k}|^{\gamma}$. The inset corresponds to the defect proliferation regime, and peak anti-hyperuniformity occurs at $\tilde \zeta_c=0.022$ and increases with $L$. (b) Sample average of $\Delta \overline N (R) = (\overline{N(R)^2} - \overline{N}(R)^2)^{1/2}$ against $\overline{N}(R)$ for several activities and radii. Giant number fluctuations with $\beta \approx 0.7 \Rightarrow \gamma \approx -0.8$ are observed for activities near the transition, whereas for $\tilde \zeta \geq 0.025$, the expected $\beta = 1/2$ scaling of uniform systems is observed.}
    \label{fig:alpha_superfig}
\end{figure*}

To quantify the spatial extent of correlations, we estimate the correlation length $\xi(\tilde \zeta)$ (averaged over realizations) as the smallest distance $r^*$ satisfying $|\overline{h}(\tilde \zeta, r^*)| - \textup{err}(\overline{h}(\tilde \zeta, r^*)) < 10^{-2}$, \sg{where $\textup{err}(\overline{h}(\tilde \zeta, r))$ is the uncertainty on $\overline{h}(r)$ for activity $\tilde \zeta$} (Fig. \ref{fig:pcf_corr_len}b). The correlation length exhibits a pronounced peak for $\tilde \zeta \in [0.0215,0.022]$.  Importantly, the correlation length grows with system size in the proliferation regime, signaling long-ranged pair correlations that are often (but not always) associated with critical transitions \cite{torquato_hyperuniform_2018, scheffer_2009}. At higher activities, however, $\xi(\tilde \zeta)$  becomes independent of system size, consistent with short-range correlations dominating in this regime.

These results indicate that the transition at $\tilde \zeta_c \approx 0.022$ is marked by the emergence of long-ranged pair correlations, with a correlation length that diverges with system size, further underscoring the critical nature of the transition.

\subsection{Emergence of Anti-Hyperuniform Defect States near Transition} \label{subsec:anti-hyperuniformity}

\noindent
A long-ranged total pair correlation function is inherently associated with hyperfluctuations and anti-hyperuniformity, motivating us to examine the spatial behavior of defect density fluctuations at long range. While density fluctuations have long been known to encode crucial systemic information about many-body systems \cite{torquato_hyperuniform_2018}, recent work—particularly the seminal contributions of Torquato and Stillinger~\cite{torquato_2003}—has highlighted the importance of their asymptotic scaling behavior in characterizing systemic properties.

Based on this scaling, systems of particles can be classified according to their asymptotic density fluctuation behavior as either \textit{uniform}, \textit{hyperuniform} or \textit{anti-hyperuniform}.
Most disordered states of matter, like e.g. ordinary fluids and amorphous solids, are uniform, meaning that $\sigma_N^2(R)$, the variance in the number of particles contained in a randomly placed spherical observation window with radius $R$, scales like $\sigma_N^2(R) \sim R^{d-\gamma}$, where $\gamma = 0$, in the limit of large $R$. For disordered hyperuniform systems, $1 > \gamma > 0$, whereas $-d \leq \gamma < 0$ for anti-hyperuniform systems \cite{torquato_hyperuniform_2018}.

Interestingly, while equilibrium examples of anti-hyperuniformity are limited to systems at thermal critical points~\cite{torquato_hyperuniform_2018}, many active systems exhibit this property as a manifestation of `giant number fluctuations'~\cite{vicsek_2012}.
In this context, the relation $\sigma_N^2(R) \sim R^{d-\gamma}$ is recast as $\sqrt{\sigma_N^2(R)} = \langle N(R) \rangle ^\beta$, where $\beta = \half(1 - \frac{\gamma}{d})$  \cite{toner_long–range_1995}.
For uniform systems, $\beta = \half$, as predicted by the law of large numbers \cite{toner_2019}, whereas $\gamma < 0$ corresponds to $\beta > \half$, with the implication that the error on the sample mean $\overline{N}(R)$ improves more slowly than $1/\sqrt{N_{\textup{samples}}}$.

The scaling behavior of density fluctuations can also be characterized through the asymptotic behavior of the structure factor $S$ or the total pair correlation function $h$. In particular, for disordered hyperuniform and anti-hyperuniform systems, $S(\vb{k}) \sim |\vb{k}|^\gamma$, $\sigma_N^2(R) \sim R^{d-\gamma}$, and $h(\vb{r}) \sim |\vb{r}|^{-(d+\gamma)}$ are equivalent in the infinite volume limit \cite{torquato_hyperuniform_2018}. 
The presence of a long-ranged pair correlation function, therefore, naturally suggests the need to investigate the scaling behavior of defect density fluctuations. 

To determine the scaling exponent $\gamma$, we fit the small-wavenumber tail of the structure factor (averaged over samples and orientations), $\overline{S}(|\vb{k}|) \sim |\vb{k}|^{\gamma}$ (see App~\ref{methods:sfac}). Evidently, defect configurations are anti-hyperuniform in the proliferation regime, with peak anti-hyperuniformity occurring at $\tilde \zeta_c = 0.022$. Notably, the strength of anti-hyperuniformity at the critical activity increases with increasing system size $L$ (Fig. \ref{fig:alpha_superfig}a). As the expression used to calculate the structure factor (Eq. \eqref{eq:methods:sfac}) becomes an increasingly accurate estimator of the true (infinite volume) structure factor as $L \rightarrow \infty$, the system size dependence of $\gamma$ is a strong indication that the defect configurations \textit{are} indeed anti-hyperuniform. 

Conversely, the seeming hyperuniformity observed at higher activities, most notably for the smallest system size, diminishes as the system size is increased. This indicates that the corresponding suppression of density fluctuations are prominent for distances below some threshold, but do not persist at long range.

To directly measure defect density fluctuations, we estimate the moments of $\overline{N}(R_i)$ using spherical observation windows for 50 radii $R_i$ linearly spaced in $[L/100, L/10]$ (see App. \ref{methods:num_var} for more details). The \sg{large-R} scaling of $\Delta \overline{N}(R) \equiv\left(\overline{\sigma_N^2}(R)\right)^{1/2}$ reveals that giant number fluctuations of topological defects are indeed present close to $\tilde \zeta_c = 0.022$, with $\beta(\tilde \zeta_c) \approx 0.7 \Rightarrow \gamma(\tilde \zeta_c) \approx -0.8$ (Fig. \ref{fig:alpha_superfig}b), in agreement with the structure factor estimate (Fig. \ref{fig:alpha_superfig}a). 
As the activity is increased further, $\beta$ gradually decreases to $\beta = 1/2$, and the defect density fluctuations transition to the $\overline{N}^{1/2}$ scaling characteristic of uniform systems.

Taken together, these findings establish that the transition at 
$\tilde \zeta_c = 0.022$ is marked by a peak in anti-hyperuniformity, with defect density fluctuations displaying giant number fluctuations and long-range correlations. Importantly, 
the observation that anti-hyperuniformity becomes more pronounced with increasing system size confirm that these anti-hyperuniform states
are intrinsic to the system as opposed to mere artifacts of finite-size effects. This underscores the critical role of activity in governing not only the density but also the spatial organization and fluctuation dynamics of topological defects in active nematics.

\vspace{1em}
\noindent
To further characterize the transition, we set out to estimate the anomalous dimension $\eta$, which determines the infinite-volume scaling of the pair correlation function through $h(r, \tilde \zeta_c) \sim r^{-(d-2+\eta)}$ at the critical activity $\tilde \zeta_c$.
Since the asymptotic scaling behavior of $h$ is also determined by $\gamma$ through $h(r) \sim {r}^{-(d+\gamma)}$, 
it follows that $\eta = d + \gamma_{c, \infty}$, where $\gamma_{c, \infty} \equiv \gamma(\tilde \zeta_c, L\rightarrow\infty)$.

\begin{figure}[h]
    \centering
\includegraphics[width=1\linewidth]{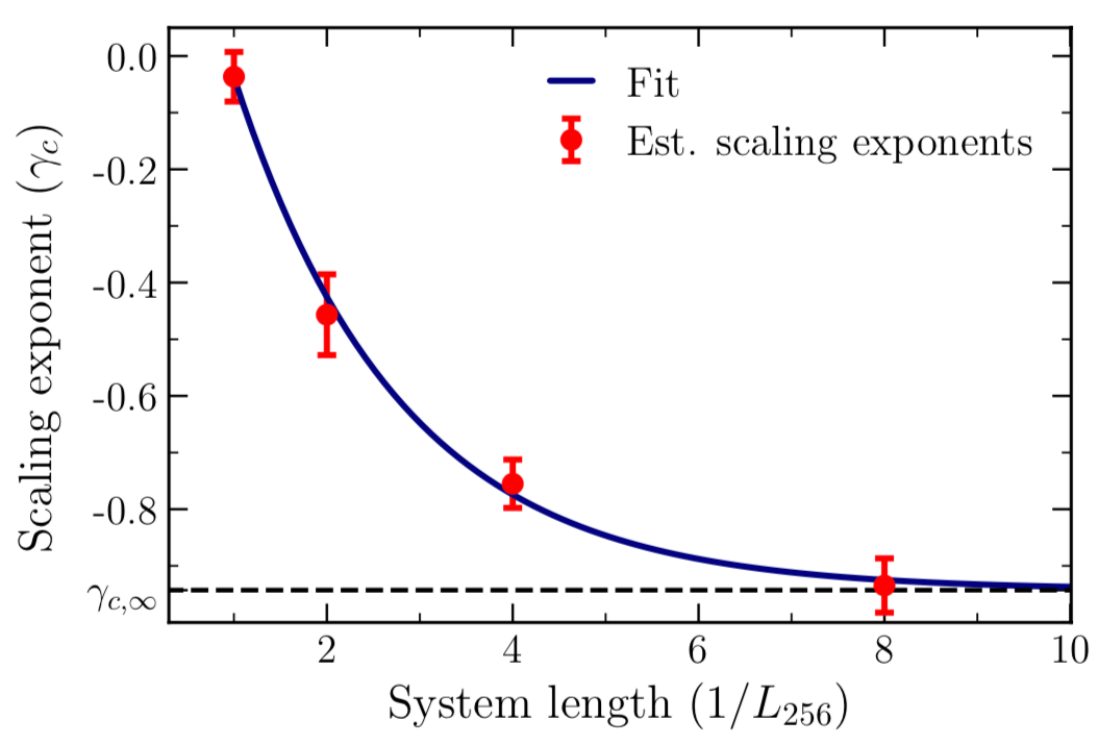}
\caption{{\bf Estimating the anomalous dimension.} 
Scaling exponents $\gamma_c(L) \equiv \gamma(\tilde \zeta=0.022,L)$ (Fig. \ref{fig:alpha_superfig}a), as a function of $L$,
along with a shifted exponential fit, $\hat\gamma_c(L) =  \gamma_c(L_{256}) + \gamma_{c,\infty} [ 1 - \exp(-\kappa (L-L_{256}))]$,  where $\kappa L_{256} = 0.56(16)$, and the estimated asymptotic scaling exponent is $\gamma_{c,\infty} = -0.94(8)$. We obtain an estimate for the anomalous dimension as $\eta = 2 + \gamma_{c,\infty} = 1.04(8)$}
\label{fig:alpha_scaling}
\end{figure}

By assuming that $\tilde \zeta = 0.022$ is sufficiently close to $\tilde \zeta_c$ for this relation to be valid, we obtain an estimate for $\gamma_{c, \infty}$ by fitting the structure factor estimates $\gamma_c(L) \equiv \gamma(\tilde \zeta=0.022,L)$ from Fig. \ref{fig:alpha_superfig}a against $L$.
Their trend is well-described by a two-parameter shifted exponential fit, $\hat\gamma_c(L) =  \gamma_c(L_{256}) + \gamma_{c,\infty} [ 1 - \exp(-\kappa (L-L_{256}))]$, where $\kappa L_{256} = 0.56(16)$, and  $\gamma_{c,\infty} = -0.94(8)$ (Fig. \ref{fig:alpha_scaling}).

Consequently, we arrive at an estimate for the anomalous dimension as $\eta = 2 + \gamma_{c,\infty} = 1.04(8)$. 
While the authors are aware of no fitting universality class characterized by this value of the anomalous dimension, and a complete characterization of the full set of critical exponents and universality class is beyond the scope of the current work, this result is a first step towards that goal.

\begin{figure*}[t!]
    \centering
\includegraphics[width=1\linewidth]{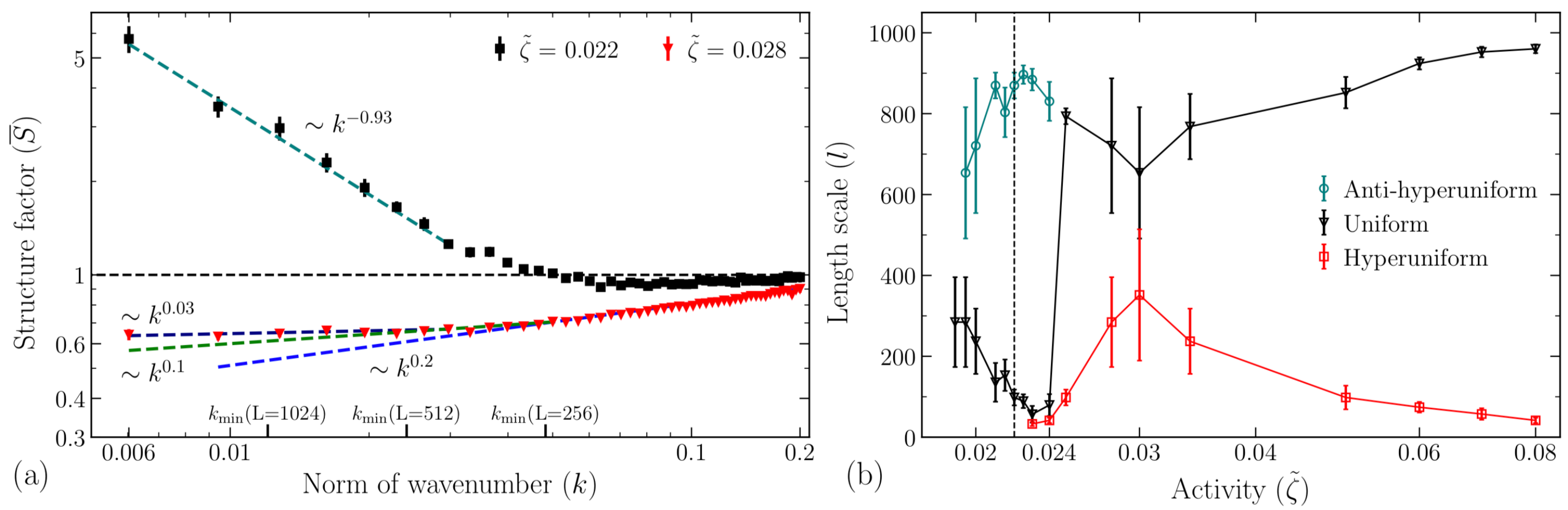}
 \caption{{\bf Active topological defects show multi-scale density fluctuations.} (a) Structure factor $\overline{S}$ (averaged over time and realizations) for $\tilde \zeta \in \{0.022, 0.028 \}$ and $L=2048$. The smallest possible wavenumber $k_\textup{min}$ is indicated for $L \in \{256,~ 512, ~1024\}$, and the dotted line at $S = 1$ indicates the theoretical value of the structure factor for the Poisson process. The asymptotic scaling of $\overline{S}$ for $\tilde \zeta = 0.022$ ($\tilde \zeta = 0.028$) is indicated by the light blue (gray) line. The blue line shows the scaling of $\overline{S}$ for $|\vb{k}| \gtrapprox k_\textup{min}(L = 256)$, and the green line the scaling for $k_\textup{min}(L = 512) \lessapprox |\vb{k}| \lessapprox k_\textup{min}(L = 256)$. (b) Estimated length scales $l = 2 \pi (k_\textup{lower}^{-1} - k_\textup{upper}^{-1})$ for $L=2048$, corresponding to the estimated wavenumber ranges $[k_\textup{lower}, k_\textup{upper}]$, in which the scaling of $\overline{S}$ is anti-hyperuniform ($\gamma < 0$), uniform ($\gamma \approx 0$), and hyperuniform $(\gamma > 0)$, respectively. The transition activity, $\tilde \zeta_c=0.022$, is indicated by a vertical line. The length scale of anti-hyperuniformity (hyperuniformity) peaks at $\tilde \zeta\in[0.021,0.024]$ ($\tilde \zeta \in [0.028,0.034]$), and the sudden drop in the uniform length scale at $\tilde \zeta = 0.024$ corresponds to the onset of anti-hyperuniformity.}
    \label{fig:sfac_superfig}
\end{figure*}

\subsection{Active Topological Defects Show Multi-Scale Density Fluctuations}

\noindent
To better understand the nature of the variations in the hyperuniformity  exponent $\gamma$, we examine the scaling behavior of the time-averaged structure factor $\overline{S}$ across the range of accessible wavenumbers (shown in Fig. \ref{fig:sfac_superfig}a for $\tilde \zeta \in \{0.022, 0.028 \}$ and $L = 2048$).
For smaller system sizes, the scaling behavior of $\overline{S}$ is qualitatively contained within the behavior observed for $L=2048$. For instance, at $\tilde \zeta = 0.028$, $\overline{S} \sim k^{0.2}$ for $|\vb{k}| > k_{\textup{min}}(L=256)$. This amounts to only estimating $\overline{S}$ on subregions smaller than $256 \times 256$, and the corresponding scaling exponent matches that found for $L=256$ (Fig. \ref{fig:alpha_superfig}a). Similarly, for $k_{\textup{min}}(L={512}) < |\vb{k}| < k_{\textup{min}}(L={256})$, the scaling $\overline{S} \sim k^{0.1}$ is consistent with the estimated scaling exponent for $512$, and so on. 

This behavior emphasizes the necessity of using sufficiently large system sizes for analyzing long-range density fluctuations in active systems. 
If one were to only consider the scaling of $\overline{S}$ for $|\vb{k}| > k_{\textup{min}}(L={256})$, for instance, one would incorrectly conclude that defects were uniform (or hyperuniform) for $\tilde \zeta = 0.022$ (or $0.028$).

At $\tilde \zeta=0.022$, the scaling behavior of the structure factor crosses over from being uniform at small length scales to anti-hyperuniform at larger length scales. This crossover behavior is characteristic of all activities where anti-hyperuniformity is observed ($\tilde \zeta \le 0.024$). Similarly, at $\tilde \zeta=0.028$, the scaling behavior of the structure factor crosses over from suppressed density fluctuations at small length scales to uniform behavior at large scales, a feature consistently observed at larger activities.

To quantify these uniformity-anti-hyperuniformity and hyperuniformity-uniformity crossovers, we calculate the length scale $l = 2 \pi (k_\textup{lower}^{-1} - k_\textup{upper}^{-1})$ corresponding to the estimated wavenumber ranges $[k_\textup{lower}, k_\textup{upper}]$, in which the scaling of $\overline{S}$ is anti-hyperuniform ($\gamma < 0$), uniform ($\gamma \approx 0$), and hyperuniform $(\gamma > 0)$, respectively (Fig. \ref{fig:sfac_superfig}b). \tm{The minimum $k_\textup{lower}$ corresponds to the minimal accessible wavenumber $2\pi/(L/2)$,
as set by the largest resolvable length scale of $L/2$ when computing the rotationally averaged structure factor.}
It is evident that anti-hyperuniformity emerges for $\tilde \zeta \leq 0.024$, and that the length scale of anti-hyperuniformity peaks for $\tilde \zeta \in [0.021,0.024]$, thus coinciding with the transition activity and that of maximal anti-hyperuniformity (Fig. \ref{fig:alpha_superfig}a). Interestingly, the length scale of uniformity drops abruptly as the activity is decreased below $\tilde \zeta=0.025$. This crossover at $\tilde \zeta \approx 0.024$ corresponds to a narrow activity range where all three length scales (uniformity, hyperuniformity, and anti-hyperuniformity) coexist. Qualitatively, this suggests that the length scale of uniformity is replaced by anti-hyperuniformity, while simultaneously replacing hyperuniformity.

The length scale of hyperuniformity peaks in 
the activity range $\tilde \zeta\in[0.028, 0.034]$, which roughly correspond to the activities for which $\gamma > 0$ is maximized (Fig. \ref{fig:alpha_superfig}a).
As the activity increases further, this length scale decreases because the onset of uniformity shifts to smaller distances. Eventually, defect density fluctuations approach the ideal gas limit at all length scales, with the upper bound determined by $L/2$.

To explain the suppression of density fluctuations at small scales and their activity dependence, we consider the defect density and the \sg{characteristic} size of ordered nematic domains, which scales with the active length 
$r_a \sim \sqrt{K/\zeta}$. Defects, \sg{being singularities in the director field, are restricted to the boundaries/interfaces between such local domains of high nematic order.}

In the limit of high activity, the density of topological defects increases \ad{without bound}, and the size of locally ordered regions \ad{diminishes}. \ad{In this regime, the chaotic flows driven by active stresses dominate over elastic relaxation, and defect motion is primarily governed by advection in the active flow field rather than by mutual elastic interactions.} As a result, positional correlations between defects are suppressed, and their spatial distribution \ad{approaches that of an ideal gas, i.e. an uncorrelated ensemble with uniform density fluctuations across all length scales}.

As the activity is lowered, defect density drops, and the size of nematic regions increases.
If the density is large enough that defects occur regularly throughout the system, and most nematic regions are surrounded by defects, the inaccessibility of ordered regions effectively restrict defect positions to a `pseudo grid' along the boundaries of such regions.
This naturally suppresses density fluctuations as compared to the uniform case, where points can occur anywhere, provided that the length scale is large enough that the observation window contains several ordered regions.

These findings highlight the rich, multi-scale nature of defect density fluctuations in active nematics, where the interplay of defect density, domain size, and activity governs transitions between anti-hyperuniformity, uniformity, and hyperuniformity. The transition activity $\tilde \zeta_c\approx 0.022$ emerges as a critical point associated with maximal anti-hyperuniformity, offering a clear marker for the onset of large-scale organization in these systems. 

\ad{Furthermore, it is important to note that the existence of anti-hyperuniform critical states and multiscale defect density fluctuations, are robust to changing boundary conditions to no-slip and free-slip boundaries (Fig. \ref{fig:suppfig:sdens_alpha_boundaries}), the inclusion of hydrodynamic screening through frictional damping (Figs \ref{fig:suppfig:dens_alpha_fric_full} \& \ref{fig:suppfig:dens_av_density_alpha_fric10_comp}), varying the flow-aligning parameter (Fig. \ref{fig:suppfig:dens_alpha_lambda}), the elastic constant (Fig. \ref{fig:suppfig:dens_alpha_elastic}), and even when $+1/2$ and $-1/2$ defects are analyzed separately (Figs. \ref{fig:suppfig:pcf_signed} \& \ref{fig:suppfigsfac_signed}).}

\begin{figure*}[!tb]
    \centering
    \includegraphics[width=\linewidth]{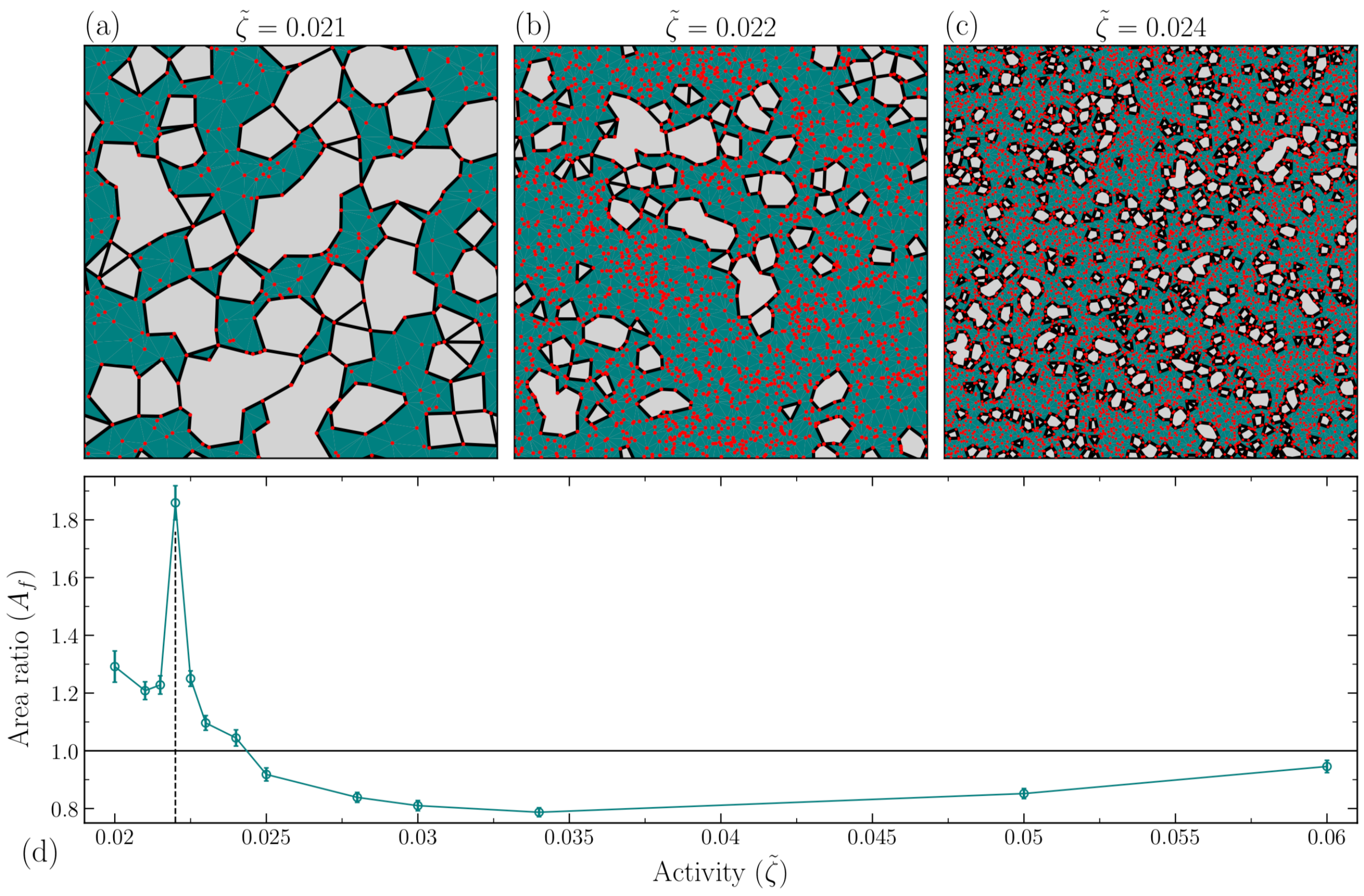}
    \caption{{\bf Emergent clustering of topological defects near transition.} (a)-(c) Examples of simulation frames with defect-free regions highlighted in gray, defects in red, and defect-laden regions in teal. The regions are identified as described in the text. (d) The ratio, $A_f =\overline{A_{\textup{nem}}^{\text{max}}} / \overline{A_{\textup{uni}}^{\text{max}}}$, between the (sample-averaged) area of the largest defect-free region in our alpha shape complex, $\overline{A_{\textup{nem}}^{\text{max}}}$, to the same result obtained for an analogous system of uncorrelated and uniformly distributed points with the same density, $\overline{A_{\textup{uni}}^{\text{max}}}$. The decline indicates a trend from existence of large defect-free region to more uniform geometries, as activity increases. Additionally, at the critical activity, $\tilde \zeta_c = 0.022$ (vertical line), we note that the ratio spikes, indicating that fluctuations stabilize defect-free regions.}
    \label{fig:PHResults}
\end{figure*}

\subsection{Emergent Clustering of Topological Defects Near Transition} \label{sec:PH}
\noindent
As anti-hyperuniformity and strong irregularity in the spatial distribution of points are often associated, we set out to examine the clustering tendency of defects. 
Specifically, we probe the geometric configuration of the defects in the nematic texture using an adapted version of persistent homology methods~\cite{Verri1993, Robins1999, Edelsbrunner2002} (see  App.~\ref{sec:PHAppendix} for details).

For a given frame, we locate the defect-free patches (Fig. 6a-c), identify the largest, and record its area, $A_{\textup{nem}}^{\text{max}}$ (see Fig. 7 for details). Finally, for each activity, we compare the sample average of the largest area, $\overline{A_{\textup{nem}}^{\text{max}}}$, 
to that obtained from the same analysis on a uniformly distributed point cloud with the same density, $\overline{A_{\textup{uni}}^{\text{max}}}$, and present the ratio of their means, $A_f$.

This area ratio peaks exactly at the critical activity $\tilde \zeta_c = 0.022$, (Fig.~\ref{fig:PHResults}d). This confirms that the average area of the largest defect-free region is significantly larger than if defects had been uniformly distributed.
The emergence of large defect-free patches near the critical activity is in line with the observed giant number fluctuations in the defect density (Fig. \ref{fig:alpha_superfig}b)

Above the transition, the area ratio drops to below $1$, indicating that the average area of the largest defect-free region becomes smaller than if defects had been uniformly distributed.
This is consistent with the results of the previous section (Fig. \ref{fig:sfac_superfig}b), namely, that defect density fluctuations are suppressed below a distance threshold in this activity regime.
As the activity is increased further, the area fraction approaches unity, reflecting that defect configurations tend toward the geometry of a uniform distribution. This approach to uniformity is consistent with our previous observation that defect density fluctuations approach uniformity at all length scales in the limit of high activity (Fig. \ref{fig:sfac_superfig}b).




This result provides statistical confirmation of the significant clustering tendency of defects for $\tilde \zeta \leq \tilde \zeta_c$, which is also readily apparent from visual inspection of the defect configurations (see supplementary videos 4-6). 
As is also evident from these videos, defect-free regions of considerable size can persist over extended periods of time,  illustrating the surprising observation that near the transition, large defect-free and defect-dense regions can coexist at steady state. This is interesting, as such a stable coexistence of defect-free regions interleaved with regions of defect aggregates has no analog in topological phase transitions in passive systems, and suggests the emergence of an intermediary biphasic state in between defect-free and defect-laden active turbulence.

\ad{A possible mechanism for this behavior is a nonlinear instability in the defect dynamics that promotes local amplification of defect density. Recent theoretical work~\cite{lavi2025} has shown that active nematics can exhibit bistability between flowing and quaiescent states due to subcritical behavior in the onset of chaotic flows. While the analysis in that work is restricted to homogeneous configurations that disallow the creation of topological defects, our full simulations suggest that the system may spatially realize this bistability by dynamically partitioning into stable, defect-suppressed regions and unstable, defect-rich regions. In this vein, our findings offer evidence that such a spatial coexistence of dynamical states can emerge through nonlinear interactions in the active nematic: regions with initially elevated defect density may further destabilize the local director field, promoting continued defect proliferation, while neighboring regions may remain or become defect-suppressed and relatively stable. This self-amplifying behavior provides a natural mechanism for the spontaneous formation of spatially inhomogeneous defect distributions and the resulting anti-hyperuniformity.}
\begin{figure*}[t!]
    \centering
    \includegraphics[width=\linewidth]{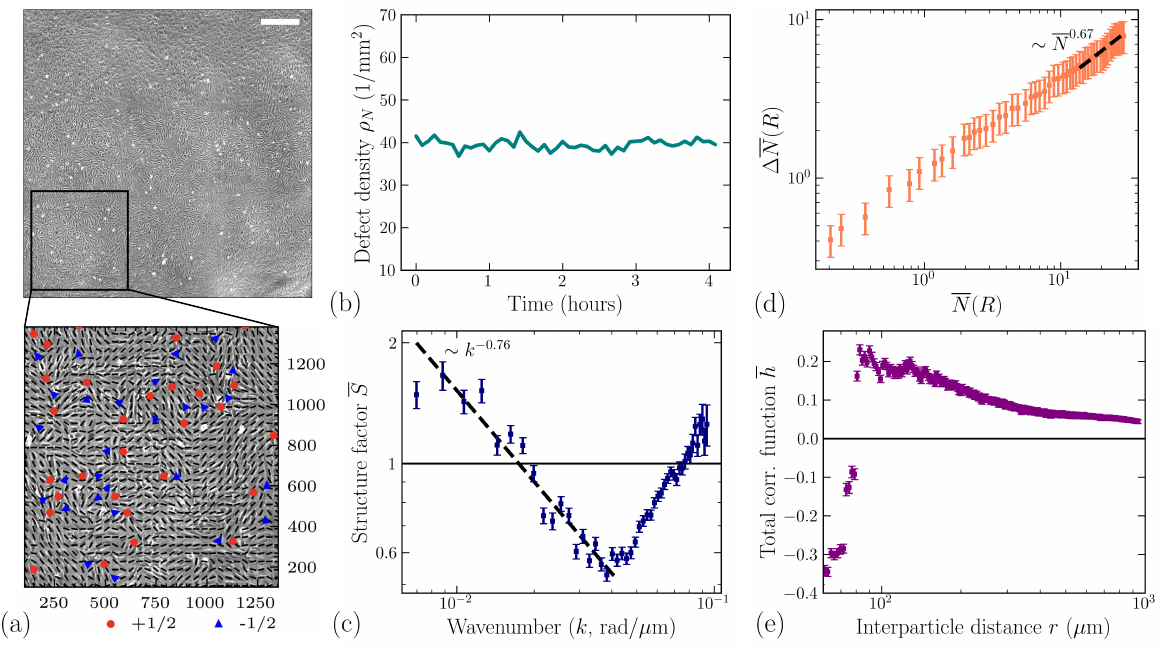}
    \caption{{\bf Evidence of anti-hyperuniformity in large-field-of-view endothelial monolayers.}
    \tm{(a) Representative phase-contrast image of an endothelial monolayer (top; scale bar: 500~$\mu$m), with a zoomed-in region (bottom) showing the local nematic director field (black line segments) and topological defects (red circles) overlaid on the image. The axes in the zoomed-in panel are labeled in $\mu$m. 
    (b) Time evolution of the global defect density $\rho_N(t)$ remains approximately constant throughout the experimental time window. 
    (c) The structure factor $\overline{S}(k)$ (averaged over time and orientations) reveals anti-hyperuniform scaling at mesoscopic scales ($\overline{S}(\mathbf{k}) \sim |\mathbf{k}|^{\gamma_S}$) with $\gamma_S = -0.76(5)$,
    with an apparent crossover to hyperuniform-like scaling at shorter scales. 
    (d) Time-averaged $\Delta \overline N (R) = (\overline{N(R)^2} - \overline{N}(R)^2)^{1/2}$ as a function of $\overline{N}(R)$ reveal giant number fluctuations consistent with anti-hyperuniformity, characterized by an exponent $\beta =0.67(3)$, corresponding to $\gamma_N = -0.68(12)$.
    (e) Time-averaged total pair correlation function $\overline{h}$ of topological defects exhibits long-range positive correlations, indicative of enhanced large-scale fluctuations consistent with anti-hyperuniformity.}}
    \label{fig:expResults}
\end{figure*}

\subsection{Experimental Evidence of Anti-hyperuniformity in Endothelial Monolayers} \label{subsec:experiments}

\noindent

\tm{To experimentally probe the main theoretical predictions on anti-hyperuniformity and multiscale defect fluctations, we analyze the spatial organization of topological defects in confluent endothelial monolayers imaged over millimeter-scale fields of view (see App.\ref{methods:expMethods} for details). The endothelial monolayer provides a robust biological realization of active nematics: elongated cell shapes define a nematic director field, and active stresses from actomyosin contractility naturally generate topological defects~\cite{ruider2024topological, keshavanarayana2024mechanical}. Previous studies have shown that the active nematohydrodynamic framework with extensile activity, employed in our simulations, closely matches defect dynamics in these systems~\cite{ruider2024topological}.}

\tm{We have performed long-term phase-contrast imaging and extract nematic director fields from local cell elongation. Topological defects are identified via computations of the winding number of from the director field~\cite{thijssen_binding_2020} (Fig.~\ref{fig:expResults}a). Over the 4-hour imaging window, the global defect density remains approximately constant (Fig.~\ref{fig:expResults}b), indicating that the system operates in a statistical steady state.}

\tm{To quantify the spatial organization of defects, we compute the structure factor $\overline{S}$ (Eq.~\eqref{eq:methods:sfac}). At mesoscopic scales, i.e., for $|\mathbf{k}| \lessapprox 4 \times 10^{-2}$, corresponding to distances larger than $50\pi ~\mu \textup{m}$ $\overline{S}$ displays clear power-law scaling, $\overline{S}(\mathbf{k}) \sim |\mathbf{k}|^{\gamma_S}$ with exponent $\gamma_S = -0.76(5)$ (Fig.~\ref{fig:expResults}c). This is consistent with anti-hyperuniform behavior ($0 > \gamma_S > -2$). Notably, at higher wavenumbers (shorter length scales), the structure factor exhibits a positive slope---indicative of hyperuniform-like suppression of fluctuations---consistent with the multi-scale crossover behavior predicted in simulations.}

\tm{To further validate the presence of anti-hyperuniformity at large scales, we calculate the number fluctuations of defects $\Delta \overline{N}(R)$ in the experimental system, following the same approach used in our simulations (App. \ref{methods:num_var}). For large window sizes $R$, fluctuations scale as $\Delta \overline{N}(R) \sim \overline{N}(R)^{0.67 \pm 0.03}$ (Fig.~\ref{fig:expResults}d).
This corresponds to a scaling exponent $\gamma_N = -0.68(12)$, which is consistent with the structure factor estimate $\gamma_S =-0.76(5)$.
Finally, the total pair correlation function $\overline{h}(r)$ reveals long-range positive correlations extending across hundreds of microns (Fig.~\ref{fig:expResults}e), supporting the presence of spatial defect clustering and long-range interactions, providing direct evidence of the spatially correlated defect organization that underlies anti-hyperuniform scaling behavior.}

\tm{These experimental measurements confirm that the anti-hyperuniformity and multiscale scaling identified in our simulations are robust, physically observable features of large-scale active nematic systems, rather than artifacts of computation or model-specific assumptions.}

\section{Discussion} \label{sec:discussion}

\noindent
Through large-scale numerical simulations of active nematics, we have uncovered compelling evidence for an active topological phase transition, which separates distinct states of defect distributions and is marked by critical dynamics of the defect organization. 
Below the critical threshold, the defect density exhibits slowed relaxation and temporal correlations, along with fluctuations characterized by large amplitudes and long lifetimes. As the activity is increased beyond the critical point, these properties grow more pronounced and culminate at the critical activity. This point marks the transition to defect-laden turbulence, in which defects are uniform, near-critical signatures of defect behavior vanish, and pair correlations between defects are negative and short-ranged.
At and below the critical activity, pair correlations become positive and long-ranged, mostly so at the transition, with correlation lengths that grow with system size. Defect configurations also exhibit anti-hyperuniformity, associated with giant number fluctuations and clustering, further emphasizing the collective organization of defects at criticality.



The nature of the phase transition and the origin of defect anti-hyperuniformity and clustering warrant further discussion. 
For equilibrium and non-equilibrium systems alike, anti-hyperuniformity is accompanied by cluster formation, long-range spatial correlations, temporal correlations, and critical slowdown \cite{torquato_hyperuniform_2018, binney_1992, marchetti_hydrodynamics_2013}. 
In active matter, anti-hyperuniformity has been observed in ordered phases of substrate-bound systems, such as polar and nematic fluids \cite{toner_long–range_1995, vicsek_novel_1995, toner_flocks_1998, ngo_gnf_2014, palacci_gnf_2013, shankar_gnf_2018, alaimo_gnf_2016}. For systems of anisotropic particles, these hyperfluctuations are attributed to the spontaneous breaking of rotational symmetry \cite{toner_2005_review, marchetti_hydrodynamics_2013}. Interestingly, anti-hyperuniformity has also been observed in systems of confined active particles without alignment interactions \cite{fily_2012,cates_2012}.
In such cases, anti-hyperuniformity emerges during phase separation, where particles in an isotropic state condense into a high-density phase \cite{marchetti_hydrodynamics_2013, vicsek_2012}. In contrast, our results show no evidence of defect ordering or the coexistence of distinct isotropic and dense defect phases. Instead, the observed anti-hyperuniformity is a direct consequence of defect clustering.



Unlike systems of active particles, where clustering arises from particle aggregation or local velocity slowdown, defects in active nematics are excitations in the director field and can be created and annihilated. This allows for a distinct nucleation mechanism where defect creation is more likely in the vicinity of existing defects, as has been shown experimentally~\cite{nieves_preprint_2024}. Such clustering dynamics could drive the emergence of anti-hyperuniformity at the transition. Indeed, while experiments \ad{on microtubule-kinesin motor mixtures~\cite{nieves_preprint_2024} and on swarming {\it B. subtilis} bacteria~\cite{yashunsky2024topological}} suggest that \ad{separate species ($+1/2$ or $-1/2$) of} topological defects in active nematics may be hyperuniform, our study reveals that \ad{total} defect density fluctuations are suppressed only for high activities and short length scales, while matching the behavior of a uniform point distribution at larger scales. 
Conversely, at lower activities, these fluctuations are uniform at small length scales but transition to anti-hyperuniformity at larger scales. This dual scaling behavior underscores the interplay between defect interactions and the active length scales dictated by nematic order.

The defect phase transition found in this study is distinct from the defect-unbinding transitions observed in equilibrium or screened active systems. In dry passive nematics, defect-unbinding coincides with the Berezinskii-Kosterlitz-Thouless (BKT) transition, driven by thermal fluctuations \cite{stein_kosterlitz–thouless_1978}. For active nematics with screening mechanisms, unbinding occurs at a critical activity, destabilizing nematic order and driving a nematic-isotropic transition \cite{shankar_defectunbinding_2018, shankar_hydrodynamics_2019}. In unconfined active nematics, however, any nonzero activity leads to instability, defect unbinding, and rapid nematic breakdown \cite{thampi_instabilities_2014, thampi_active_2016}. Our findings reveal a critical transition beyond the unbinding regime, where defects are already unbound, highlighting its fundamentally different origin. Importantly, the signatures of critical behavior are observed across system sizes, underscoring the robustness and significance of the transition. To build on these findings, a full characterization of the transition and its universality class is essential.

Finally, our results offer new insights into the transition from defect-free to defect-laden active turbulence.
This transition appears mediated by a biphasic state, where defect-free and defect-dense regions coexist at steady state. Defects, being self-propelled and sources of vorticity \cite{thampi_2014}, actively influence the structure and dynamics of the system. The coexistence of these regimes suggests a gradual restructuring, with defect clustering playing a central role in the emergence of defect-laden turbulence. These findings highlight the intricate collective dynamics of active defects, as well as their role in organizing the transition to defect-laden turbulence. To refine our understanding of this process, further work is needed to examine the transition in terms of fluid flows, vorticity, and related properties.

In conclusion: By connecting defect clustering, anti-hyperuniformity, and criticality in defect organization, these findings reveal new insights into the rich physics of active matter, as well as the distinct states and generic critical behavior of active topological defects.

\section*{Data availability statement}
\noindent
The data that support the findings of this study are available
upon reasonable request to the authors.

\section*{Acknowledgements}
\noindent
S. G. A. thanks Lasse Frederik Bonn and Jayeeta Chattopadhyay for enlightening discussions in relation to this work, as well as Lasse Frederik Bonn, Carl Gustav Henning Hansen, Francesco Ferrarin, Daria Gusew and Maria Gabriela Jordão Oliveira for their valuable feedback on the readability and clarity of the manuscript.
M. C. P. thanks Villum Fonden for financial support (grant no. VIL69081). A. D. acknowledges funding from the Novo Nordisk Foundation (grant no. NNF18SA0035142 and NERD grant no. NNF21OC0068687), Villum Fonden (grant no. 29476), and the European Union (ERC, PhysCoMeT, 101041418). Views and opinions expressed are however those of the authors only and do not necessarily reflect those of the European Union or the European Research Council. Neither the European Union nor the granting authority can be held responsible for them.

\appendix
\section*{Appendix} \label{sec:appendix}

\subsection{The Hybrid Lattice Boltzmann Method}
\label{methods:hb}

\noindent
The governing Eqs. \eqref{eq : gov-equations} are solved using the hybrid Lattice Boltzmann method developed by Marenduzzo et al. \cite{Marenduzzo_hlb_2007}. The evolution of the flow velocity $u_i$ is determined by solving the generalized Navier-Stokes Eqs. (\eqref{eq:stokes_cont} and \eqref{eq:stokes_full}) using the Lattice Boltzmann method (LBM), while the time evolution of nematic order parameter $Q_{ij}$ is found by solving the Beris-Edwards Eq. \eqref{eq:beris-ed} using a
five-point stencil approximation. For the LBM, we have used a $D2Q9$ grid for discretizing the velocity, and the so-called BKG collision operator to model the collision term in the Boltzmann equation \cite{bhatnagar_bkg1954}, with a relaxation time equal to the simulation time step.

\subsection{Detecting Defects}
\label{methods:detecting_defects}

\noindent
Defects are found by calculating the winding number of the director, specifically, by employing the algorithm by Zapotocky et al. as follows \cite{zapotocky_kinetics_1995}: For every square of $2\times2$ neighboring lattice points in the system, the cumulative change in the orientation of the director is calculated as it moves around the square in the counterclockwise direction. A directional change of $\pi$ ($-\pi$) indicates the presence of a $+1/2$ ($-1/2$) defect at the center of the square.

\subsection{Estimating the Structure Factor}
\label{methods:sfac}

\noindent
As is common when working with long-range density fluctuations of points configurations, we assume the spatial distribution of topological defects at steady state to be generated by a translationally invariant and ergodic point process \cite{torquato_hyperuniform_2018}.

The former assumption entails that the pair correlation function $g_2$ satisfies $g_2(\vb{r}_1 + \vb{y},\vb{r}_2 + \vb{y})= g_2(\vb{r}_1,\vb{r}_2)$ for any $\vb{y} \in \mathbb{R}^d$, and thus $g_2(\vb{r}_1,\vb{r}_2) = g_2(\vb{r})$, where $\vb{r}=\vb{r}_2 - \vb{r}_1$. With this assumption, the structure factor in the infinite volume limit is given by $S(\vb{k}) = 1 + \rho_I \int_{\mathbb{R}^2} d\vb{r} ~ (g_2(\vb{r}) - 1) e^{-i\vb{k}\cdot \vb{r}}$, where $\rho_I$ is the number density in the infinite volume limit \cite{torquato_hyperuniform_2018}.
Assuming ergodicity implies that any realization of the ensemble is representative of the ensemble in the infinite volume limit, so that volume averages in this limit equal the corresponding ensemble averages \cite{torquato_hyperuniform_2018}. In particular, we can approximate $\rho_I$ by the ensemble average of the number density $\langle \rho_N \rangle$.

On a square domain with side length $L$, Hawat et al. and Rajala et al. have shown that the infinite-volume expression for $S$ given above can be estimated by \cite{Hawat_2023, rajala2022}
\begin{align}
    \hat{S} ( \vb{k}) = \frac{1}{\langle \rho_N \rangle L^2} \left| \sum_{j=1}^N e^{-i \vb{k} \cdot \vb{x}_j} \right|^2, ~~~~ \vb{k} \in \mathbb{A}_L,
       \label{eq:methods:sfac}
\end{align}
where $\vb{x}_j$ refers to the point positions, $\mathbb{A}_L$ denotes the set of allowed wavenumbers, i.e. the set of $\vb{k}$ for which each component $k_i = 2\pi n /L$ for $n \in \left\{1,2,...\right\}$. This estimate is asymptotically unbiased, in that the ensemble average $\langle \hat{S} ( \vb{k}) \rangle $ converges to the true structure factor $S(\vb{k})$ as $L \rightarrow \infty$. 

\subsection{Estimating the Local Defect Density and its Moments}
\label{methods:num_var}

\noindent
For each frame, we count the number of defects within a randomly placed spherical window centered at $\vb{x_0} \in [0, L] \times  [0, L]$ with radius $R_i$, and for computational efficiency, the center $\vb{x_0}$ is kept fixed while varying $R_i$ from $R_{\textup{min}}$ to $R_{\textup{max}}$ under periodic boundary conditions. 
It is well-established that choosing observation windows not much smaller than the system size leads to an under-estimation of the local number variance \cite{torquato_2021, román_1999}. The corresponding error term is proportional to $S(0)$, which makes anti-hyperuniform systems particularly prone to such bias if $R_{\textup{max}}$ is chosen too large. 

We have found that choosing $R_{\textup{max}} = L/10$ leads to no discernible bias and thus consider 50 window sizes $R_i$ linearly spaced in $[L/100, L/10]$.
Having found the number of points within a circle with radius $R_i$ for each frame, the average number of points and higher moments are estimated empirically.

\subsection{Estimating the Pair Correlation Function}
\label{methods:pcf}

\noindent
We assume the pair correlation function, $g_2$, to be rotationally invariant, so that $g_2(\vb{r}) = g_2(r)$. On a square domain with side length $L$, an estimate for the pair correlation, $\hat{g}_2$, is obtained as $\hat{g}_2(r) = \hat{K}'(r)/(2\pi r)$, where $\hat{K}$ is an estimator of the Ripley function given by
\begin{align}
    \hat{K}(r) = \frac{L^2}{N(N-1)} \sum_{i,j \neq i} \mathds{1}(d_{ij} \leq r) e_{ij} (r),
    \label{eq:ripley}
\end{align}
where the sum is taken over all ordered pairs of points, $N$ is the total number of points, $\mathds{1}(d_{ij}\leq r) = 1$ if the distance $d_{ij} \leq r$ and 0 otherwise, and $e_{ij}$ are the edge correction weights \cite{ripley_1988}.
We use Ripley's isotropic edge correction weights, which counteract bias from estimating the summand in Eq. \ref{eq:ripley} for points closer than $r$ to the boundary of the system by assuming rotational invariance of the underlying point process (for more details, see \cite{ripley_1977, ohser_1983}).

Given estimates $\hat{K}(r)$, we obtain $\hat{g}_2(r)$ by applying smoothing to $Y(r) = \hat{K}(r)/(2 \pi r)$ subject to $Y(0)=0$ and estimating $Y'(r)$. To minimize bias from the estimation of $ \hat{K}(r)$ on point configurations exhibiting clustering, $r_{\textup{max}}$ should be chosen as $r_{\textup{max}} \leq L/4$ \cite{ripley_1977}, and we calculate $\hat{g}_2$ for $512$ values of $r \in [0, r_{\textup{max}} = L/4]$.

\subsection{Simulation Parameters}
\label{methods:sim_params}

\noindent
Each simulation has been initialized in a nematic state with director angle $\theta_0$, which is then perturbed at each lattice point $\vb{x}$ with noise $n_0$ as $\theta(\vb{x}, t = 0) = \theta_0(\vb{x}) + n_0 U(-\pi, \pi)$, where $U$ is the uniform distribution.

\begin{table}[H]
\begin{tabular}{llll}
\hline
\small
Parameter                           & Symbol    & Value     & Dimension (2D)     \\ \hline \hline
Activity                 & $\zeta$                         & $[0.019,0.1]$      & $M/T^2$         \\
Alignment parameter            & $\lambda$ & $1$       & $1$           \\
Density                        & $\rho$    & $100$     & $M/L^2$       \\
Frank elastic constant         & $K$       & $0.05$    & $ML^2/T^2$       \\
Isotropic viscosity            & $\eta$    & $1$       & $M/T$         \\
Landau coefficient & $\mathcal{A}$       & $1$      & $M/T^2$       \\
Rotational diffusivity         & $\Gamma$    & $0.05$    & $T/M$         \\
Friction & $\mu$    & $0$    & $M/(L^2T)$         \\
Initial noise in alignment    & $n_0$                  & $0.05$                        & 1         \\
Initial director orientation    & $\theta_0$                  & $0$                        & 1         \\
\hline
\end{tabular}
\caption{Values/ranges and dimension of simulation parameters. $L, M, T$ refer to length, mass and time, respectively. All parameters are expressed in units of ($\Delta x$, $\Delta t$, $\mathcal{A}$).}
\label{tab:model_params}
\end{table}

\subsection{Determination of Defect-Free Regions and their Area}\label{sec:PHAppendix}

\begin{figure*}[!tb]
    \centering
    \includegraphics[width=\linewidth]{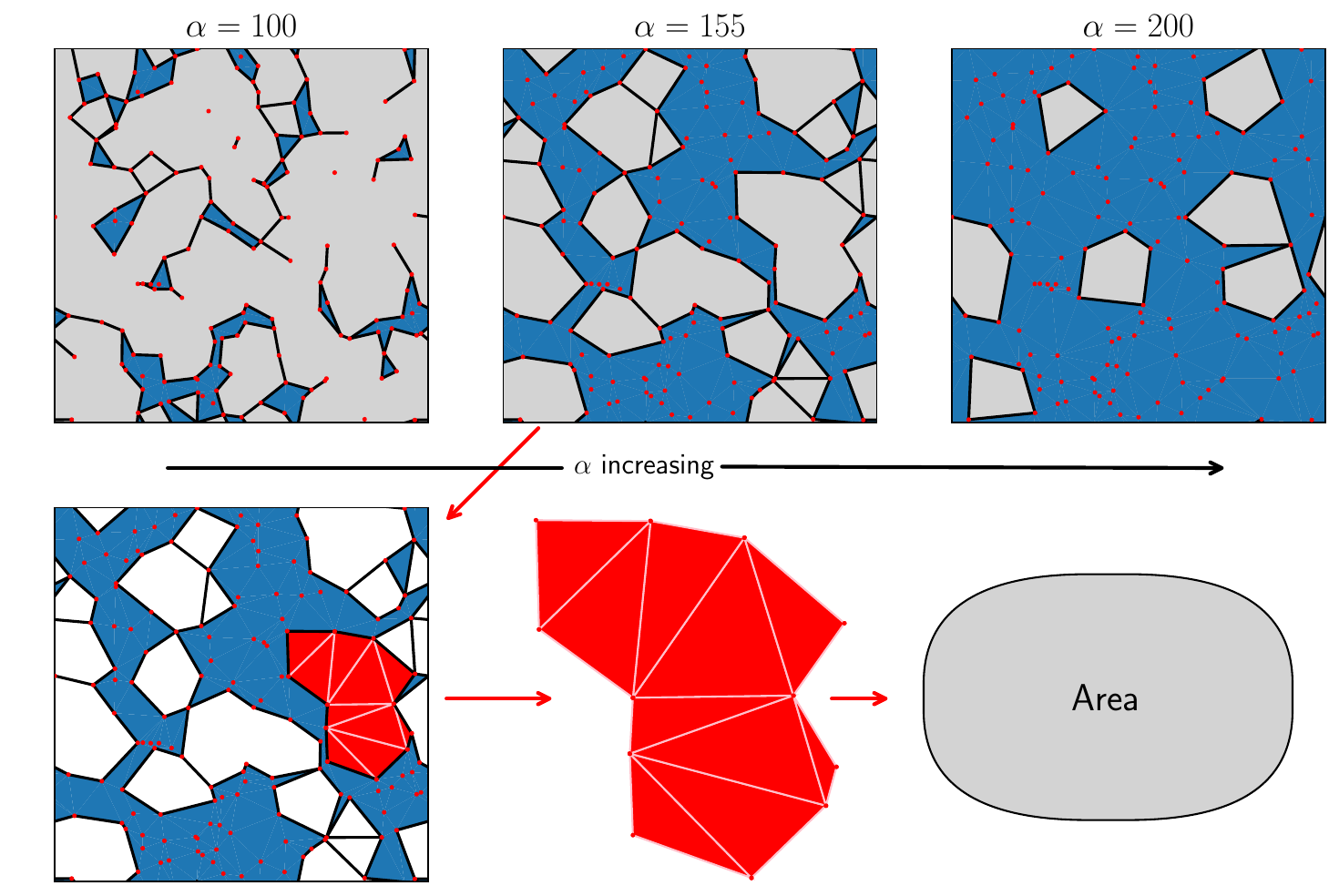}
    \caption{Above, we see three simplicial complexes in an alpha shape filtration of a given frame from our simulations with the same color scheme as Figure~\ref{fig:PHResults}. For increasing alpha, we observe how more and more of the complex is `filled in', and in particular how all defects \tm{(in red)} can eventually be found in one of the faces of the complex \tm{(in blue)}. In this case, this happens for $\alpha \approx 155$ depicted in the center. \tm{Below this value, some defects are only connected to other defects via edges (in black) or not at all (consider e.g. the isolated defects in the gray areas for $\alpha = 100$).} We continue our analysis on this frame, for which we identify the largest defect-free patch and compute its area.}
    \label{fig:PHExplanation}
\end{figure*}

\noindent
We select a subset of our simulation frames for further analysis; chosen at random from frames for which the number of defects in the simulation has stabilized. We sample $50$ such frames for $\tilde \zeta < 0.023$; and repeat this sampling for $5$ realizations of our simulations, and $10$ frames from $10$ realizations for $\tilde \zeta \geq 0.023$.

For each of these frames, we compute (periodic) alpha shape filtrations from the location of the defects in the nematic texture~\cite{Edelsbrunner1995, Edelsbrunner2010}. This yields a series of topological spaces constructed from the union of balls centered at the defect positions with gradually increasing assigned radii (see Figure~\ref{fig:PHExplanation} for $3$ examples of our alpha shape complexes). These are computed using CGAL~\cite{CGAL2025}. The literature holds several examples of similar characterizations of the geometry of amorphous and crystalline materials~\cite{Hiraoka2016, Lee2017, Pedersen2020} as well as soft~\cite{Smith2024, Pedersen2024} and granular matter~\cite{Saadatfar2017}. In this series, we identify the first member for which all defects are contained in faces of the given alpha shape complex and focus our remaining analysis on this length scale and geometry. We then compute the areas of all patches remaining in the complex and identify the largest (as shown in Figure~\ref{fig:PHExplanation}). In practice, we calculate these by computing the Delaunay triangulation~\cite{Delaunay1934} of the defect geometry, identifying the faces contributing to our defect-free region by their circumradius, and appropriately summarizing the areas of these.

For comparison, for a given simulation frame we generate $5$ uniformly distributed point clouds with the same density, compute alpha shape filtrations of these as well, and - as for our simulation - identify the minimal value for alpha for which all the remaining patches are defect-free and compute the same areas of defect-free regions (see Figure~\ref{fig:PHResults}).

\subsection{Endothelial Cell Culture and Imaging}
\label{methods:expMethods}

\noindent
\tm{Primary human umbilical vein endothelial cells (HUVECs) were isolated from newborn umbilical cords as described previously~\cite{ma2024coupling,su2024microfluidic}. Informed consent was obtained from all donors, and all procedures were approved by the Beihang University Ethics Committee. Cells were cultured in endothelial cell medium (ECM; ScienCell) supplemented with 5\% fetal bovine serum (FBS), 1\% endothelial cell growth supplement (ECGS), and 1\% penicillin--streptomycin, and maintained at 37\,\textdegree{}C in a humidified atmosphere with 5\% CO\textsubscript{2}. Cells between passages 2 and 6 were used for all experiments.}

\tm{Endothelial monolayers were established on polydimethylsiloxane (PDMS; Sylgard 184, Dow Corning) films supported on glass substrates. The PDMS films, approximately 20 $\mu$m in thickness, were prepared by spin-coating, with agent ratio and curing temperature carefully controlled to yield mechanical properties comparable to physiological stiffness \cite{su2024microfluidic}. Prior to seeding, PDMS surfaces were sterilized under ultraviolet light for 1 hour and coated with 80\,\si{\micro\gram}/mL fibronectin (Cat: 356008, Corning) at 37\,\textdegree{}C for 1 hour. Cells were seeded and statically cultured for about 2 hours to form a confluent and uniform endothelial monolayer.}

\tm{Time-lapse imaging was performed using phase-contrast microscopy at 10$\times$ magnification (DMi8; Leica) equipped with a high-resolution CMOS camera (DFC9000 GT; Leica). To obtain a large field of view, nine adjacent subfields were imaged and stitched together to form a single composite image. Images were acquired every 5 minutes over a total duration of 12 hours. Analysis was restricted to the final 4-hour window, during which the global defect density remained stationary and the system reached a statistically steady state.}

\tm{We then used OrientationJ and followed our previous methods~\cite{saw2017topological} to extract the nematic director field from endothelial monolayers. Topological defects were subsequently identified using the same approach as in the simulations, by computing the winding number of the director field.}

\clearpage
\bibliography{active_matter_final}




\clearpage
\renewcommand{\thesection}{\arabic{section}} 
\setcounter{subsection}{0}                      
\setcounter{figure}{0}
\renewcommand{\thefigure}{S\arabic{figure}}
\section*{Supplementary Information}

\noindent
\textit{Page numbering as well as references to articles, figures, equations etc. are in relation to those of the original article}

\label{sec:supplementary_information}

\noindent
\subsection*{Analyzing $+1/2$ and $-1/2$ defects separately} \label{subsec:signed_ahu_analysis}

\begin{figure*}[!tb]
    \centering
\includegraphics[width=1\linewidth]{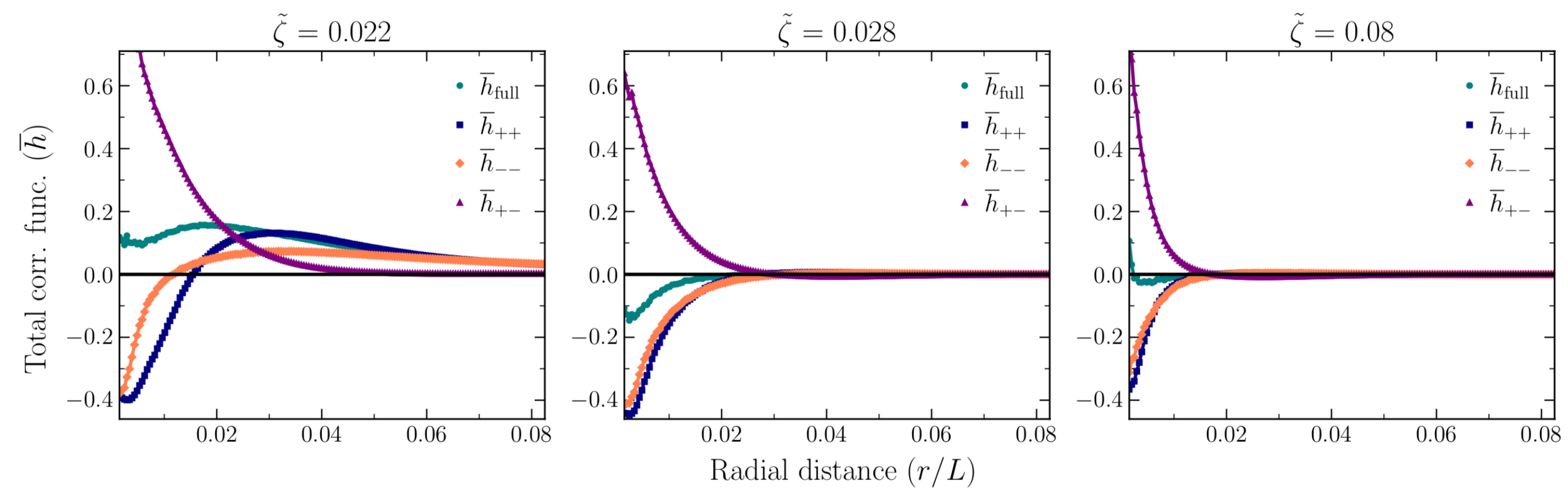}
\caption{\sg{\textbf{Total pair correlation functions accounting for defect charge for $L=2048$.} \textit{Left:} 
$\overline{h}_{++}$ and $\overline{h}_{++}$ behave differently at small distances, and the $++$ and $--$ anti-correlations transform into positive correlations as the distance is increased. Notably, long-range correlations are present even when considering $+1/2$ and $-1/2$ defects separately, and we see that $\overline{h}_{\textup{full}}  \approx \overline{h}_{++}  \approx \overline{h}_{--}$ at large distances. \textit{Center:} Strong $++$ and $--$ anti-correlations make $\overline{h}_{\textup{full}}$ negative at short distances, and all pair correlations decay quickly with distance. \textit{Right:} $\overline{h}_{\textup{full}} \approx 0$ everywhere, as expected for a uniform point distribution, and all pair correlations decay quickly with distance.}}
\label{fig:suppfig:pcf_signed}
\end{figure*}

\noindent
\sg{In the following, we set out to determine how the anti-hyperuniformity and long-ranged pair correlations of defects are affected by considering $+1/2$ and $-1/2$ defects separately. The analysis is carried out for $L=2048$ and  $\tilde \zeta \in \{0.022, 0.028, 0.08\}$. For $\tilde \zeta = 0.022$, defects are uniform at short range and become anti-hyperuniform at long range. 
For $\tilde \zeta = 0.028$, defects are hyperuniform at short range and become uniform at long range, whereas 
for $\tilde \zeta = 0.08$, defects are uniform almost everywhere (Fig. \ref{fig:sfac_superfig}).

\vspace{1em}
\noindent
To determine the role of defect charge in the long-ranged pair correlations observed  the full pair correlation function $g_{\textup{full}}$, we calculate the pair correlation functions for positive and negative defects separately ($g_{++}$ and $g_{--}$). Denoting the interaction term by $g_{+-}$, we have that $g_{+-} = 2g_{\textup{full}} - (g_{++} + g_{--})$. The corresponding total pair correlation functions, $h \equiv g-1$, can be seen in Fig. \ref{fig:suppfig:pcf_signed}.

For $\tilde \zeta = 0.08$, the pair correlation functions look exactly like we would expect for a uniform gas of defects. At short range,
the position of defects are anti-correlated with other defects of the same charge, while being positively correlated with defects of opposite charge. The full pair correlation function $\overline{h}_{\textup{full}}(r, \tilde \zeta =0.08) \approx 0$ for all $r$, as is expected for a uniform point distribution, while all other pair correlations decay quickly with distance. 

For $\tilde \zeta = 0.028$, increased anti-correlations of same-charge defects, combined with decreased correlations between defects of opposite charge, make the full correlation function negative at short range. This is consistent with the increase in nearest-neighbor distance associated with hyperuniformity \cite{nieves_preprint_2024}.

For $\tilde \zeta = 0.022$, same-sign defects are anti-correlated at short distances, but the increased $+-$ correlations make the total pair correlation function positive at short distance - in contrast to the other two cases. 
Notably, the near-symmetry in $\overline{h}_{++}$ and $\overline{h}_{--}$ at small distances is broken for $\tilde \zeta = 0.022$, and the $++$ and $--$ anti-correlations transform into positive correlations as the distance is increased.

For large distances, $\overline{h}_{\textup{full}}  \approx \overline{h}_{++}  \approx \overline{h}_{--}$, which is to say: Long-range pair correlations are present even when considering the subpopulations of positive or negative defects separately, and the long-range behavior/decay of such correlations match that of the full pair correlation function.}

\begin{figure*}[!tb]
    \centering
\includegraphics[width=1\linewidth]{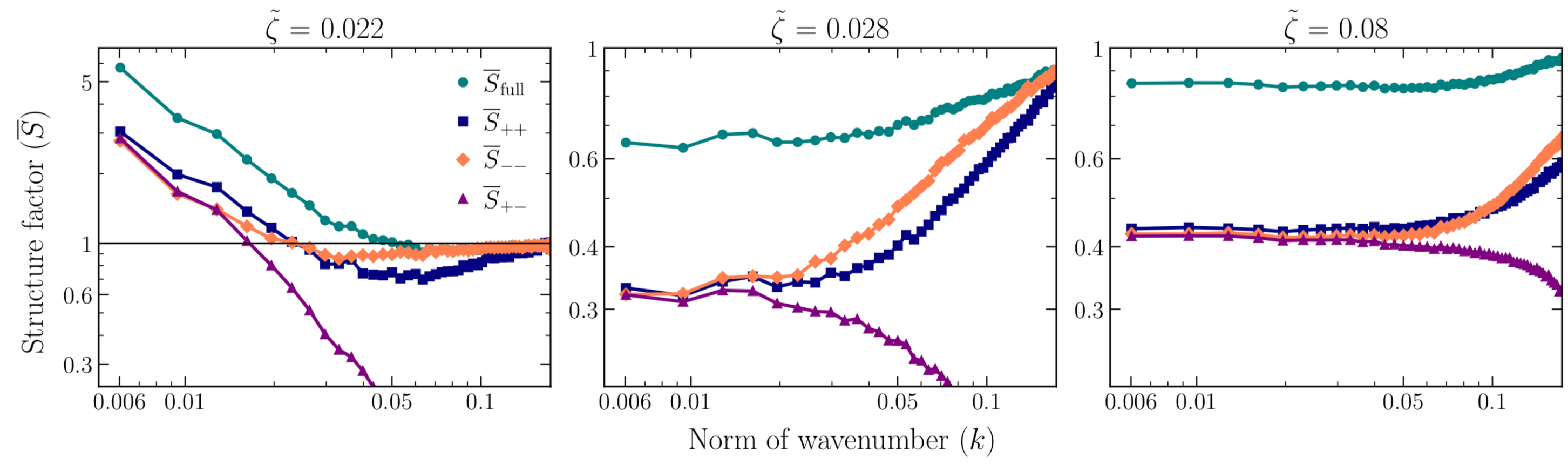}
\caption{\sg{\textbf{Structure factors accounting for defect charge for $L=2048$.}
\textit{Left:} All structure factors are increasing as $k \rightarrow 0$. Further, we see that $S_{--} \approx S_{++} \approx S_{+-} \approx \half S_{\textup{full}}$ for small $k$, 
and further that the asymptotic scalings - which determines the hyperuniformity exponents - are roughly the same in all cases. 
It follows that defect anti-hyperuniformity is observed even 
when considering $+1/2$ or $-1/2$ defects separately.
\textit{Center:} All structure factors eventually attain near-uniformity as $k$ is decreased, while for $k \geq 0.03$, $S_{++}$ and $S_{--}$ decay more quickly than $S_{\textup{full}}$. It follows that
the suppression of density fluctuations is stronger for the $-1/2$ and $+1/2$ subpopulations than for all defects in this regime.
\textit{Right:} Uniformity (as quantified by a constant slope) is quickly attained for all structure factors as $k \rightarrow 0$. 
}}
\label{fig:suppfigsfac_signed}
\end{figure*}

\vspace{1em}
\noindent
\sg{Let us now repeat the charge-sensitive analysis for the structure factor. Starting from the definition of the discrete structure factor (Eq. \eqref{eq:methods:sfac}), the interaction term can be shown to have the form $S_{+-} = S_{\textup{full}} - \half(S_{++} + S_{--})$, and the corresponding structure factors can be seen in Fig. \ref{fig:suppfigsfac_signed}. 

For $\tilde \zeta = 0.08$, uniformity (as quantified by a constant slope) is quickly attained for all structure factors as $k \rightarrow 0$. 

For $\tilde \zeta = 0.028$, near-uniformity is eventually attained as $k$ is decreased, and the small-wavenumber scaling is roughly the same for all structure factors.
We note that $S_{\textup{full}}$, $S_{--}$ and $S_{++}$ are all decreasing for $k \geq 0.03$, or equivalently for 
$L_r \leq 210$ in lattice units. This corresponds to the length regime of hyperuniformity (Fig. \ref{fig:sfac_superfig}). Restricting our attention to $k \geq 0.03$,
which amounts considering only subsystems of size $L_r \times L_r$,
we note that $S_{--}$ and $S_{++}$ decrease more quickly than $S_{\textup{full}}$. It follows that the suppression of density fluctuations is stronger for the $-1/2$ and $+1/2$ subpopulations than for all defects in this regime, consistent with the findings of Nieves et al. \cite{nieves_preprint_2024}.

Finally, for $\tilde \zeta = 0.022$, all structure factors are increasing as $k \rightarrow 0$. 
Further, we see that $S_{--} \approx S_{++} \approx S_{+-} \approx \half S_{\textup{full}}$ for small $k$, 
and similarly that the asymptotic scaling exponents are roughly the same in all cases. That is to say: The emergent defect anti-hyperuniformity is observed even when considering $+1/2$ or $-1/2$ defects separately.}

\subsection*{Additional Simulation Results}
\label{subsec:additional_simulations}

\noindent
\sg{
To probe the robustness of the transition and the activity regime of defect anti-hyperuniformity, we have carried out 
additional long-runtime ($n_{\textup{timesteps}} \sim 4\cdot 10^6$) simulations 
with varying boundary conditions, frictional coefficient, alignment parameter and elastic constant. In all cases, $L=512$ with four realizations per activity, and apart from the parameter being varied, all parameters are kept fixed and equal to those of the original simulations (Tab. \ref{tab:model_params}).}

\subsubsection{Varying Boundary Conditions}
\label{subsubsec:boundaries}

\begin{figure*}[!tb]
    \centering
\includegraphics[width=1\linewidth]{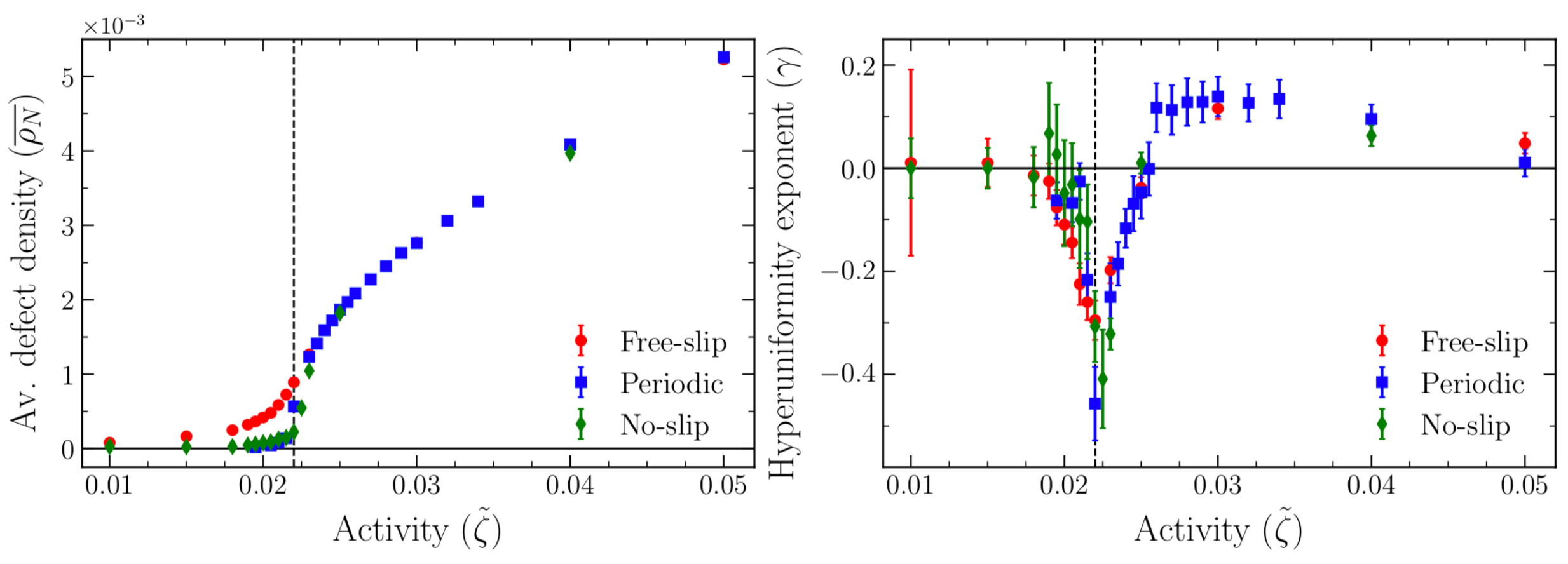}
\caption{\sg{\textbf{Effect of using 
different boundary conditions for simulations with side length $L=512$}. The dashed vertical line at $\tilde \zeta_c = 0.022$ indicates the transition activity for systems with periodic boundary conditions (as used in the article). \textit{Left}: Average global defect density $\overline{\rho_N}$ against activity. At high activities, the defect densities for each case roughly coincide. 
At small activities, periodic boundary conditions suppresses defect creation, and no defects are present for $\tilde \zeta \lessapprox 0.019$. 
In contrast, defect creation becomes favorable at smaller activities
for systems with free-slip and no-slip boundary conditions, respectively. 
For the no-slip case, $\overline{\rho_N}$ remains small and roughly constant for $\tilde \zeta \lessapprox 0.02$. For the free-slip case, it grows smoothly even at low activities, and the proliferation regime is stretched as compared to the periodic and no-slip systems. \textit{Right}: Estimated (structure factor) hyperuniformity exponents. The choice of boundary conditions does not appear to have any effect on the anti-hyperuniformity of defects and its activity dependence.}}
\label{fig:suppfig:sdens_alpha_boundaries}
\end{figure*}

\noindent
\sg{To determine the robustness of our results to non-periodic boundaries, we have replaced the periodic boundary condition (BC) with \textit{free-slip} and \textit{no-slip} BC, respectively. No-slip BC constrains the flow field velocity to 0 at the boundary, while free-slip BC constrains the component of the flow velocity normal to the boundary to 0, but leaves the parallel component unconstrained.

In Fig.~\ref{fig:suppfig:sdens_alpha_boundaries}, we compare the average defect density and the estimated hyperuniformity exponents for the free-slip and no-slip cases with our original results with periodic BC. 
While the defect densities for each case roughly coincide at high activities, periodic boundary conditions suppresses defect creation for small activities, and no defects are present for $\tilde \zeta \lessapprox 0.019$ (Fig. \ref{fig:suppfig:sdens_alpha_boundaries} left). 
For the no-slip and free-slip systems, however, defect creation becomes favorable at much smaller activities. For the no-slip case, the defect density remains small and roughly constant for $\tilde \zeta \lessapprox 0.02$. For the free-slip case, the defect density grows smoothly even at low activities, and the proliferation regime is stretched as compared to the periodic and no-slip systems. 

Despite these differences in the activity-dependence of the global defect density, the choice of boundary conditions has no significant effect on the anti-hyperuniformity of defects and its activity dependence. The strength of anti-hyperuniformity at the transition activity is possibly less for the free-slip system as compared to the periodic and no-slip systems, but this difference is not statistically significant.

In conclusion, the emergence of an anti-hyperuniform regime is unaffected by exchanging the periodic with the more physical no-slip or free-slip boundary conditions.}

\subsubsection{Adding Friction}
\label{subsubsec:friction}

\begin{figure*}[!tb]
    \centering
\includegraphics[width=1\linewidth]{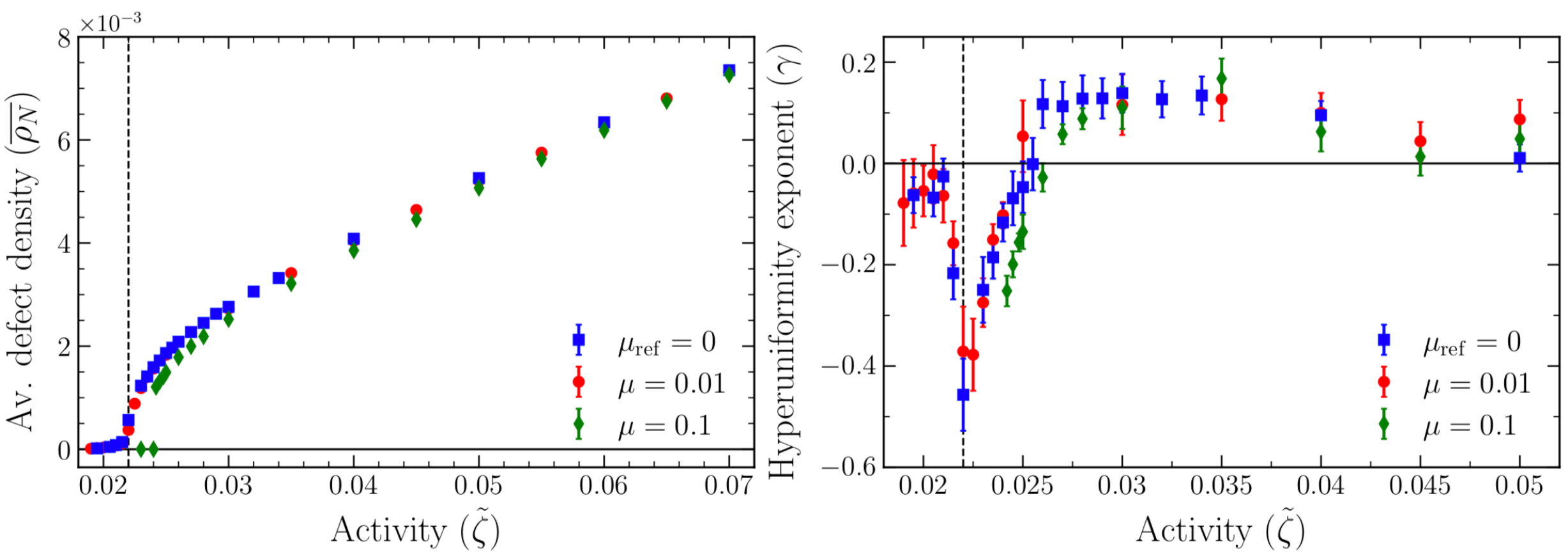}
\caption{\sg{\textbf{Effect of including frictional damping with coefficient $\mu$ for simulations with side length $L=512$}. The dashed vertical line at $\tilde \zeta_c = 0.022$ indicates the transition activity for systems with $\mu_{\textup{ref}}=0$ (as in the article). \textit{Left}: Average global defect density $\overline{\rho_N}$ against activity.
Whereas the $\mu=0.01$ case leaves $\overline{\rho_N}$ largely unaffected, 
the $\mu = 0.1$ case shows an increase in the activity threshold for defect nucleation from $\tilde \zeta \approx 0.019$ (for $\mu \leq 0.01$) to $\tilde \zeta \approx 0.024$.
\textit{Right}: Estimated (structure factor) hyperuniformity exponents. Including friction with $\mu=0.01$ does not appear to have any effect on the strength of defect anti-hyperuniformity or its activity dependence. 
For $\mu=0.1$, defects become increasingly anti-hyperuniform as $\tilde \zeta \searrow 0.024$. Below this activity, no defects are present, and the anti-hyperuniformity regime is partly truncated.}}
\label{fig:suppfig:dens_alpha_fric_full}
\end{figure*}

\begin{figure*}[!tb]
    \centering
\includegraphics[width=1\linewidth]{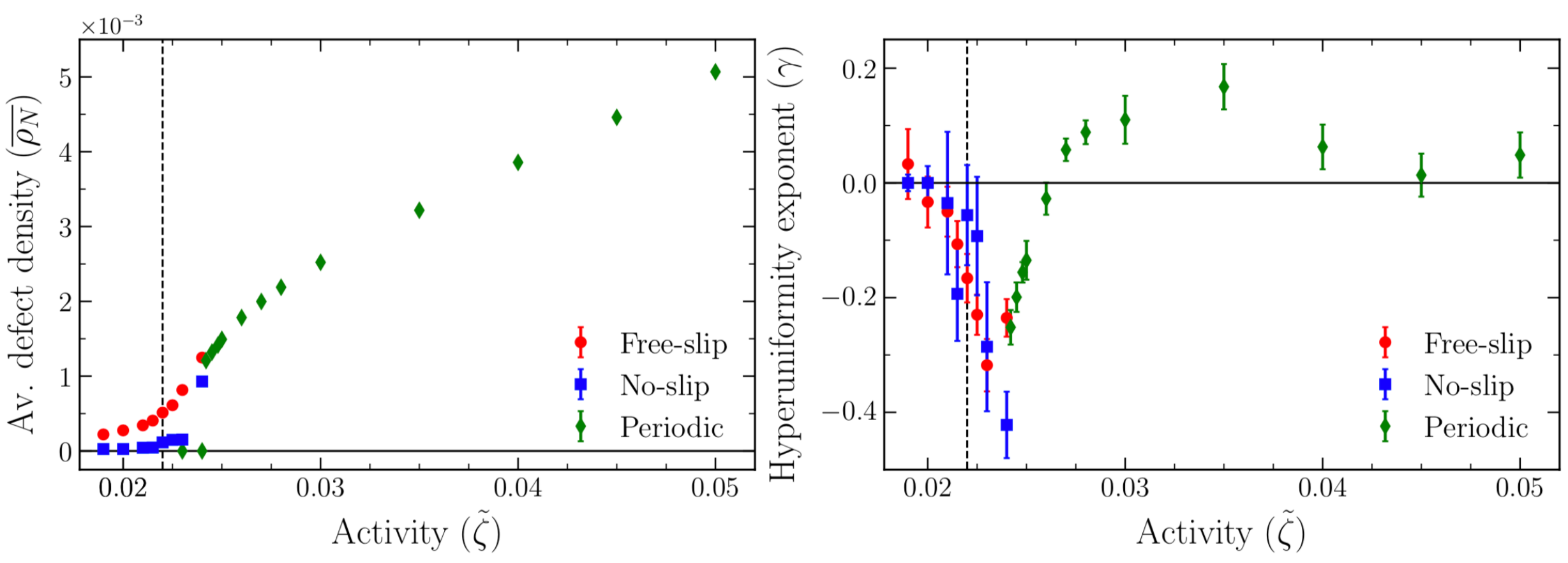}
\caption{\sg{\textbf{Effect of using 
different boundary conditions for simulations with frictional coefficient $\mu = 0.1$ and side length $L=512$}. \textit{Left}: Average global defect density $\overline{\rho_N}$ against activity. Replacing periodic with no-slip or free-slip boundaries decreases the activity threshold for defect nucleation (as already seen in Fig. \ref{fig:suppfig:sdens_alpha_boundaries} left for $\mu=0$). 
\textit{Right}: Estimated (structure factor) hyperuniformity exponents. 
With periodic boundaries, no defects are for present for $\tilde \zeta \lessapprox 0.024$, and the anti-hyperuniformity regime is truncated accordingly. 
Replacing periodic with no-slip or free-slip boundaries allows defect creation for $\tilde \zeta < 0.024$ and restores the truncated part of the anti-hyperuniformity regime.}}
\label{fig:suppfig:dens_av_density_alpha_fric10_comp}
\end{figure*}

\begin{figure}[h]
    \centering
\includegraphics[width=1\linewidth]{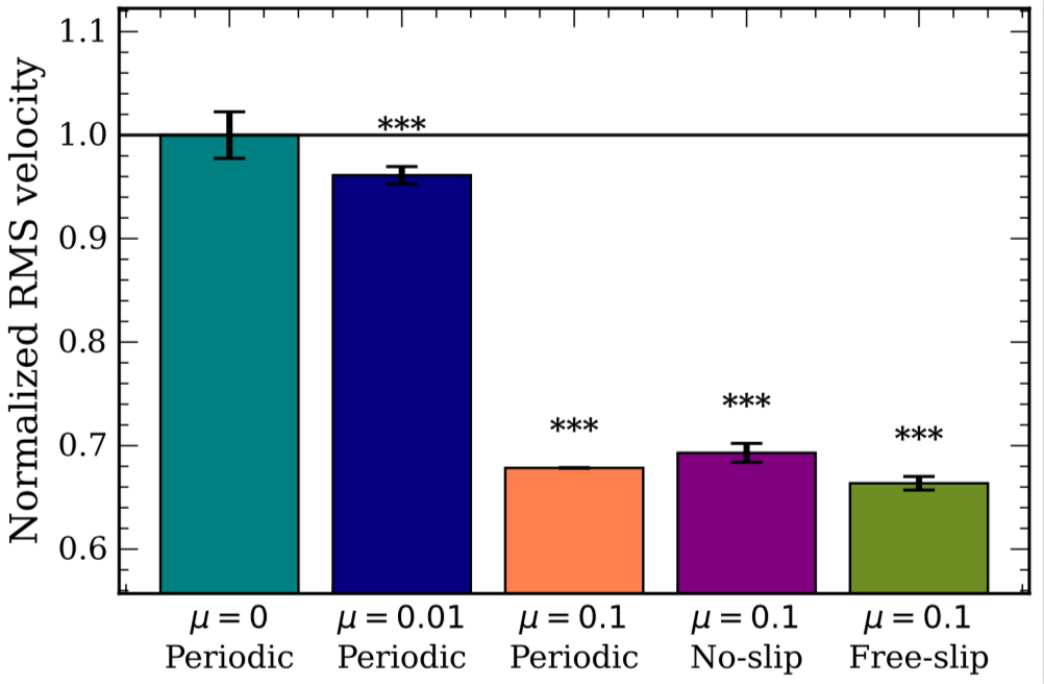}
\caption{\sg{\textbf{Effect of frictional damping and boundary conditions on system dynamics.} Root-mean-square (RMS) velocity of the flow field, normalized to the control ($\mu = 0$, periodic), is shown for different friction coefficients and boundary conditions at fixed activity $\tilde{\zeta} = 0.024$. Low friction ($\mu = 0.01$) leads to a mild reduction in motion, whereas moderate/strong friction ($\mu = 0.1$) significantly suppresses dynamics. This suppression persists when moderate/strong friction is combined with no-slip or free-slip boundary conditions. Statistical significance is assessed by two-sample $t$-tests relative to the control (***$p < 0.001$, $n = 4$, mean $\pm$ SEM).}}
\label{fig:suppfig:rms_comparison}
\end{figure}

\noindent
\sg{To determine the robustness of our results to the inclusion of frictional damping, we compare our original simulation with frictional coefficient $\mu_{\textup{ref}} = 0$ to simulations with $\mu \in \{0.01, 0.1\}$
Friction which enters into the equation of motion for the velocity field $u_i$ as $-\mu u_i$ (Eq. \eqref{eq:stokes_full}). It has units of $M/L^2T)$, and the value of $\mu$ sets the screening length through $l_{\textup{sc}} = \sqrt{\eta / \mu}$, above which frictional damping dominates over viscous dissipation \cite{hemingway_correlation_2016}. With $\eta = 1$, this is the case for $l > 10$ when $\mu = 0.01$ and $l > \sqrt{10}$ when $\mu=0.1$.

To ensure that these values of $\mu$ correspond to non-negligible frictional damping near the transition activity $\tilde \zeta_c = 0.022$, we have calculated the RMS-velocity of the flow field for each $\mu$ and $\tilde \zeta = 0.024$ (Fig. \ref{fig:suppfig:rms_comparison}).
For $\mu = 0.01$, the root-mean-square (RMS)-velocity is decreased by about $5\%$ as compared to the frictionless case, whereas for $\mu = 0.1$, the corresponding decrease is about $30 \%$ and could be considered `moderate' to `strong' friction (for a rigorous analysis on frictional damping in active nematics, see \cite{thampi_friction2014}).

Including frictional damping with coefficient $\mu=0.01$ has no significant effect on the average defect density $\overline{\rho_N}$. 
For the $\mu = 0.1$ case, $\overline{\rho_N}$ is largely unaffected at high activity, whereas the activity threshold for defect nucleation is increased from $\tilde \zeta \approx 0.019$ to $\tilde \zeta \approx 0.024$ (Fig. \ref{fig:suppfig:dens_alpha_fric_full} left). This is consistent with previous findings \cite{thampi_friction2014}.

As for anti-hyperuniformity, including friction with $\mu = 0.01$ does not appear to have any effect on the strength of defect anti-hyperuniformity or its activity dependence. 
For the $\mu = 0.1$ case, however, matters are complicated by the fact the defect creation is suppressed for $\tilde \zeta \lessapprox 0.024$. 
In this case, defects are anti-hyperuniform for $\tilde \zeta \in (0.024, 0.025]$ with increasing strength as ~$\tilde \zeta \searrow 0.024$. and so it appears that the transition region is truncated rather than shifted in this case (Fig. \ref{fig:suppfig:dens_alpha_fric_full} left).

As we have already seen that periodic boundaries suppress defect creation at low activity (Fig. \ref{fig:suppfig:sdens_alpha_boundaries} left), we
set out to determine how the truncation seen for $\mu = 0.1$ with periodic boundaries is be affected by instead using free-slip and no-slip boundaries, respectively. As the RMS velocity for the free-slip and no-slip systems are roughly similar to that of the periodic system for $\mu=0.1$, we are still in the regime of `moderate' to `strong' friction (Fig. \ref{fig:suppfig:rms_comparison}).
Indeed, exchanging periodic with free-slip or no-slip boundaries lowers the defect nucleation threshold and restores the truncated part of the anti-hyperuniformity regime, although it appears the transition activity---the activity of maximal anti-hyperuniformity---is shifted towards slightly higher activities as compared to the $\mu=0$ case (Fig. \ref{fig:suppfig:dens_av_density_alpha_fric10_comp}).

In conclusion, the emergence of an anti-hyperuniform regime is robust to the inclusion of frictional damping.}

\subsubsection{Varying the Alignment Parameter}
\label{subsubsec:alignment}

\begin{figure*}[!tb]
    \centering
\includegraphics[width=1\linewidth]{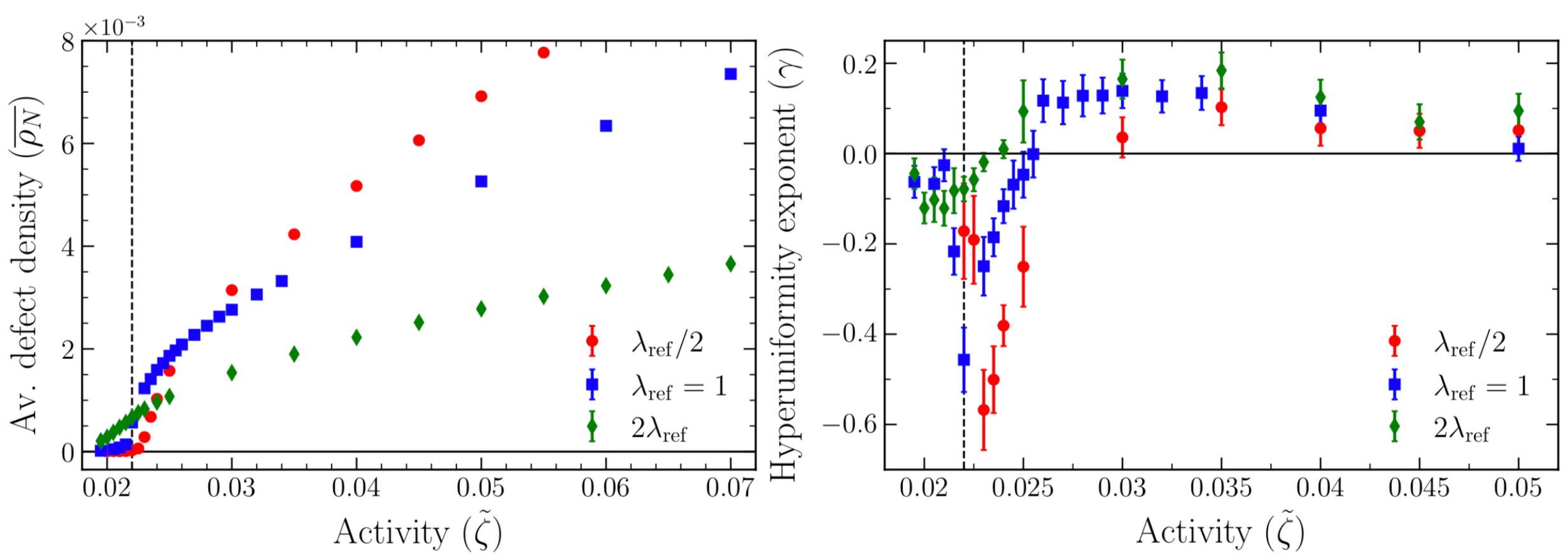}
\caption{\sg{\textbf{Effect of varying the alignment parameter, $\lambda$, for simulations with side length $L=512$}. The dashed vertical line at $\tilde \zeta_c = 0.022$ indicates the transition activity for systems with $\lambda_{\textup{ref}}=1$ (as used in the article).
\textit{Left}: Average global defect density $\overline{\rho_N}$ against activity. Decreasing $\lambda$ compresses the graph of $\overline{\rho_N}$ and
shifts the threshold for defect creation to larger activities. Conversely, increasing $\lambda$ has the opposite effect.
\textit{Right}: Estimated (structure factor) hyperuniformity exponents, $\gamma$.  Decreasing $\lambda$ shifts the regime of anti-hyperuniformity towards larger activities and seemingly increases the strength of anti-hyperuniformity at the transition. Conversely, increasing $\lambda$ shifts the regime of anti-hyperuniformity towards smaller activities and decreases the strength of anti-hyperuniformity at the transition.}}
\label{fig:suppfig:dens_alpha_lambda}
\end{figure*}

\noindent
\sg{To probe the effect of varying the alignment parameter $\lambda$, we compare our original simulation with $\lambda_{\textup{ref}}=1$ to simulations with $\lambda \in \{\lambda_{\textup{ref}}/2, 2 \lambda_{\textup{ref}}\}$. 
Choosing $\lambda = 0.5$ ($\lambda = 2$) puts the nematic well within the flow-tumbling (flow-aligning) regime \cite{hemingway_correlation_2016}.

Decreasing $\lambda$ compresses the graph of the average defect density $\overline{\rho_N}$ and shifts the regime of defect proliferation to larger activities (Fig. \ref{fig:suppfig:dens_alpha_lambda} left). Conversely, increasing $\lambda$ makes $\overline{\rho_N}$ less sensitive to changes in the activity, and the defect proliferation regime, $\tilde \zeta \in [0.02, 0.024]$, which for $\lambda=1$ is characterized by an exponentially growing defect density, has been stretched to the point that the defect density grows roughly linearly with $\tilde \zeta$.

Further, we see that letting $\lambda \rightarrow \lambda/2$ shifts the regime of anti-hyperuniformity towards larger activities and seemingly increases the strength of anti-hyperuniformity at the transition - although not significantly so (Fig. \ref{fig:suppfig:dens_alpha_lambda} right). Conversely, increasing $\lambda$ shifts the regime of anti-hyperuniformity towards smaller activities and decreases the strength of anti-hyperuniformity at the transition.

These simulations confirm that the emergence of an anti-hyperuniform regime is robust to letting $\lambda = 1 \rightarrow (0.5, 2)$ (corresponding to flow-tumbling and flow-alignment, respectively), albeit the critical activity is shifted, and the strength of anti-hyperuniformity in the transition regime is altered.}

\subsubsection{Varying the Elastic Constant}
\label{subsubsec:elastic}

\noindent
Finally, we set out to probe the effect of varying the Frank elastic constant $K$ by comparing our original simulation with $K_{\textup{ref}}=0.05$ to simulations with $K \in \{K_{\textup{ref}}/2, 2 K_{\textup{ref}}\}$. 

\begin{figure*}[!tb]
    \centering
\includegraphics[width=1\linewidth]{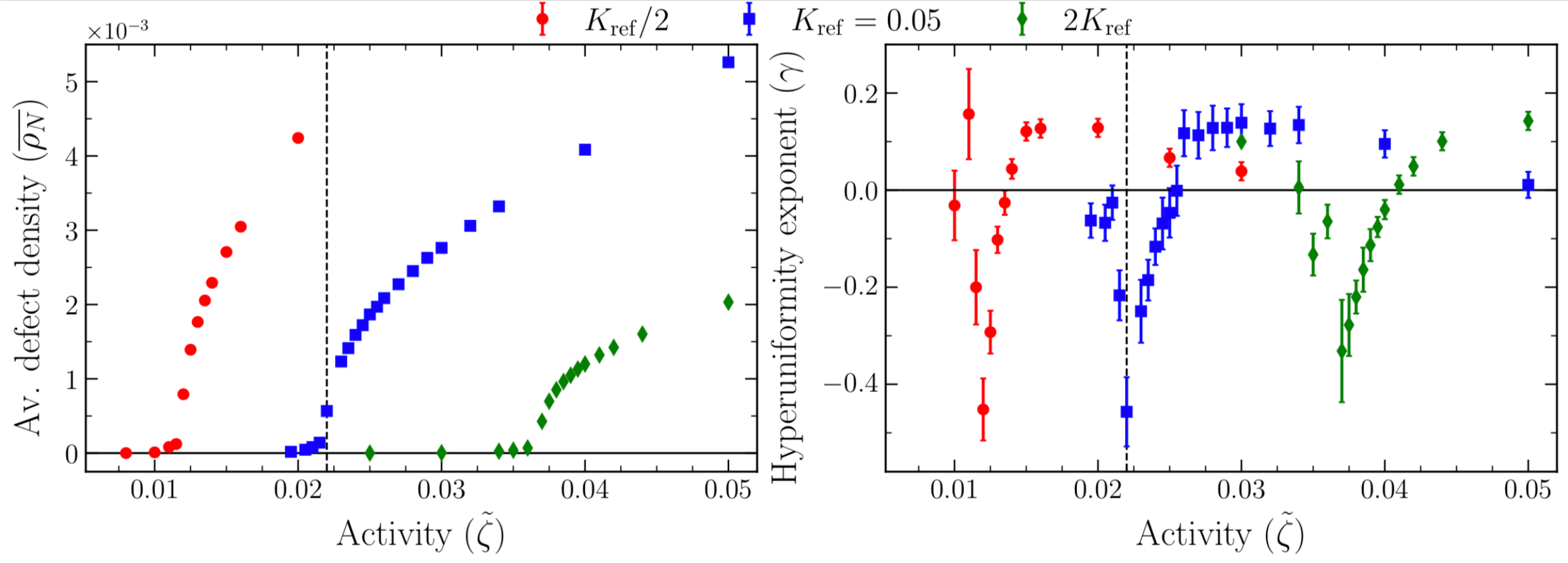}
\caption{\sg{\textbf{Effect of varying the elastic constant, $K$, for simulations with side length $L=512$.} The dashed vertical line at $\tilde \zeta_c = 0.022$ indicates the transition activity for systems with $K_{\textup{ref}}=0.05$ (as used in the article). \textit{Left}: Average global defect density $\overline{\rho_N}$ against activity. Decreasing $K$ compresses the graph of $\overline{\rho_N}$ and shifts it towards smaller activities, while increasing $K$ has the opposite effect. 
\textit{Right}: Estimated (structure factor) hyperuniformity exponents. Decreasing $K$ compresses and shifts the regime of anti-hyperuniformity towards smaller activities, while increasing $K$ has the opposite effect. The strength of anti-hyperuniformity at the transition is unchanged within statistical uncertainty.}}
\label{fig:suppfig:dens_alpha_elastic}
\end{figure*}

We see that letting $K \rightarrow K/2$ compresses the width of the anti-hyperuniformity regime and shifts the transition activity towards lower activities, while letting $K \rightarrow 2K$ has the opposite effect (Fig. \ref{fig:suppfig:dens_alpha_elastic}). In both cases, the strength of anti-hyperuniformity at the transition is statistically identical to that of the original simulations. 

These simulations confirm that the emergence of an anti-hyperuniform regime is robust to changing elastic constants, albeit the critical activity value is shifted, and the regime of anti-hyperuniformity is compressed or stretched.

\end{document}